\documentclass[iop,apj,tighten]{emulateapj}
\usepackage{apjfonts} 
\usepackage{rotating}
\usepackage{array}
\usepackage{afterpage}
\usepackage{float}
\usepackage{morefloats}
\usepackage{graphicx}
\usepackage{amsmath}
\usepackage{color}
\usepackage[breaklinks,colorlinks,citecolor=blue,linkcolor=magenta]{hyperref}

\newcommand{\OIII}{\mbox{O\,\textsc{iii}}}

\newcommand{\kms}{km s$^{-1}$}
    
\newcommand{\degree}{$^\circ$}

\slugcomment{ApJ published}

\shorttitle{The Prevalence of gas outflows in type 2 AGN\lowercase{s}. II.}
\shortauthors{Bae \& Woo}

\begin{document}

\title{The Prevalence of gas outflows in type 2 AGN\lowercase{s}. II. 3D biconical outflow models }

\author{Hyun-Jin Bae$^{1,2}$}
\author{Jong-Hak Woo$^{2,}$\altaffilmark{3}}

\affil{$^{1}$Department of Astronomy and Center for Galaxy Evolution Research, Yonsei University, Seoul 120-749, Republic of Korea; hjbae@galaxy.yonsei.ac.kr} 
\affil{$^{2}$Astronomy Program, Department of Physics and Astronomy, Seoul National University, Seoul 151-742, Republic of Korea; woo@astro.snu.ac.kr}
\altaffiltext{3}{Author to whom any correspondence should be addressed}

\begin{abstract}
We present 3D models of biconical outflows combined with a thin dust plane for investigating the physical properties of the ionized gas outflows and their effect on the observed gas kinematics in type 2 active galactic nuclei (AGNs). Using a set of input parameters,
we construct a number of models in 3D and calculate the spatially integrated velocity and velocity dispersion for each model. 
We find that three primary parameters, i.e., intrinsic velocity, bicone inclination, and the amount of dust extinction, mainly determine the simulated velocity and velocity dispersion. Velocity dispersion increases as the intrinsic velocity or the bicone inclination increases, while velocity (i.e., velocity shifts with respect to systemic velocity) increases as the amount of dust extinction increases. 
Simulated emission-line profiles well reproduce the observed [\OIII] line profiles, e.g., a narrow core and a broad wing components. By comparing model grids and Monte Carlo simulations with the observed [\OIII] velocity--velocity dispersion (VVD) distribution of $\sim$39,000 type 2 AGNs, we constrain the intrinsic velocity of gas outflows ranging from $\sim$500 \kms\ to $\sim$1000 \kms\ for the majority of AGNs, and up to $\sim$1500--2000 \kms\ for extreme cases. The Monte Carlo simulations show that the number ratio of AGNs with negative [\OIII] velocity to AGNs with positive [\OIII] velocity correlates with the outflow opening angle, suggesting that outflows with higher intrinsic velocity tend to have wider opening angles.
These results demonstrate the potential of our 3D models for studying the physical properties of gas outflows, applicable to various observations, including
spatially integrated and resolved gas kinematics.

\end{abstract}


\section{Introduction}
\label{sec:intro}
Observational studies have found scaling relations between the mass of the central black holes (BHs) and their host galaxy properties
in the local universe \citep[e.g.,][]{2000ApJ...539L...9F, 2000ApJ...539L..13G, 2009ApJ...698..198G, 2013ApJ...764..184M, 2013ARA&A..51..511K, 2013ApJ...772...49W}. Theoretical studies predict that energetic feedback from active galactic nuclei (AGNs) have played a crucial role in shaping the scaling relations \citep[see][for review]{, 2015ARA&A..53..115K}. 
The narrow-line region (NLR) of AGN is one of the best places to witness the AGN feedback as a form of ionized gas outflows based on, for example, the study of the prominent optical line [\OIII] $\lambda$5007 
\citep[e.g.,][]{2005AJ....130..381B,2008ApJ...680..926K,2011ApJ...732....9G,2011ApJ...737...71Z,2013MNRAS.430.2327L,2014ApJ...795...30B,2014MNRAS.441.3306H}. A number of spatially resolved spectroscopic studies revealed strong signatures of gas outflows driven by AGNs \citep[e.g.,][]{1994ApJ...433...48V,2000ApJ...532..247C, Veilleux:2001is, Das:2005ff,2009MNRAS.396....2B,2013ApJS..209....1F,2014ApJ...786....3S, 2014MNRAS.442..784Z,2016ApJ...819..148K}. However, understanding the intrinsic NLR kinematics from the measurements of the [\OIII] line profile, e.g., line width, is limited due to the projection effects and the complex geometry of the region. 

\begin{figure*}[t]
\centering
\includegraphics[width=0.95\textwidth]{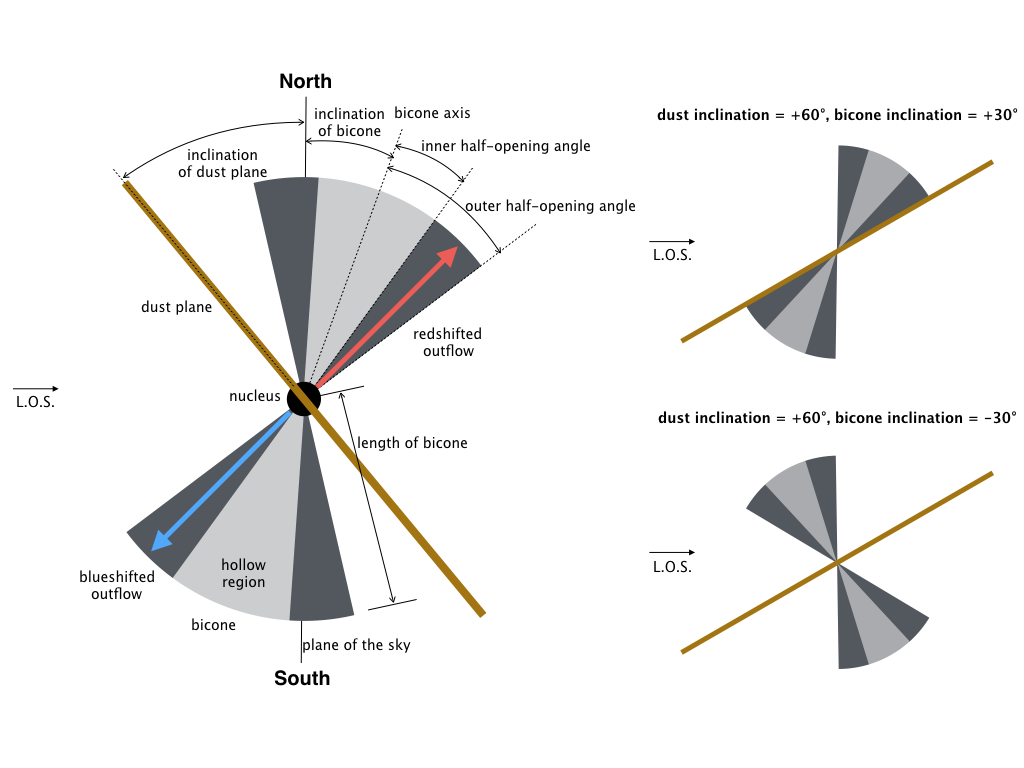}
\caption{Dissection of the bicone combined with the thin dust plane (left), and examples of the different combinations of the bicone and dust-plane orientations (right). The bicone consists of two identical cones whose apex is located at the nucleus, and the cones are axisymmetric with respect to the bicone axis. The bicone extends with an outer half-opening angle and has an inner hollow region defined by an inner half-opening angle. The radial profiles of velocity and flux can be set arbitrary (see the text for details). For a dust inclination of $+$60\degree\ (right panel), the approaching or the receding cone is obscured if the bicone inclination is $+$30\degree\ (top-right) or $-$30\degree\ (bottom-right), respectively.}
\label{fig:bicone}
\end{figure*}

A (bi-)conical morphology is one of the characteristic features of the outflows observed in local Seyfert galaxies, for example, via \textit{Hubble Space Telescope} ($HST$) narrow-band imaging \citep[e.g.,][]{1988ApJ...328..519P,1988ApJ...332..702P,1996ApJ...463..498S}. \citet{1996ApJ...463..498S} showed that such a conical shape is more frequently observed in type 2 (obscured) Seyfert than type 1 (un-obscured) Seyfert galaxies, which is consistent with the expectation from the AGN unification model \citep[e.g.,][]{1995PASP..107..803U}. 
\citet{2000ApJ...532L.101C} devised a simple 3D biconical outflow model to reproduce the position--velocity (PV) diagram of NGC 1068 measured based on the {\it HST/STIS} long-slit data. The model consists of two cones with an identical geometries, i.e., opening angle and size, with a simple profile of radial outflow velocity. By allowing the bicone to be arbitrarily oriented, they found a good agreement between the velocity constructed using the model and the observed spatial distribution of the [\OIII] velocity along the outflow direction. 

Since then, various versions of bicone models have been widely used to investigate the kinematics of the NLR in local Seyfert galaxies \citep[e.g.,][]{2000ApJ...532..247C,Veilleux:2001is,Das:2005ff,StorchiBergmann:2010fg, 2010AJ....140..577F,2011ApJ...739...69M,2013ApJS..209....1F} and quasars \citep[e.g.,][]{2014ApJ...786....3S, 2014MNRAS.442..784Z}. For example, \citet{Das:2005ff} modeled 
the NLR kinematics of a Seyfert galaxy NGC 4151, which are spatially resolved based on the $HST/STIS$ data, by modifying the radial velocity profile.
Later, \citet{StorchiBergmann:2010fg} modified the bicone model by considering weak emission from the steeply inclined wall of the cone and a constant radial velocity, in order to reproduce the PV diagram of NGC 4151 measured based on the integral-field spectroscopy obtained with the $Gemini/NIFS$. 
Also, \citet{2011ApJ...739...69M} determined the kinematics of the outflows in seven Seyfert galaxies, using biconical outflow models as well as the integral-field spectroscopy based on the $Very Large Telescope/SINFONI$. They added a rotating gas model to better reproduce the observations. \citet{2010ApJ...708..419C} combined the biconical outflow model with a thin dust plane, which can partly obscure the outflows 
from the line-of-sight (LOS), reproducing the blueshifted [\OIII] of NGC 1068 and NGC 4151 in their spatially integrated spectra.

While the bicone models have successfully reproduced the 1D (long-slit), 2D (IFU) velocity distribution, and the flux-weighed velocity from fiber or aperture spectroscopy, these models in the previous studies have several limitations. First, the parameter space, such as bicone geometry and kinematics, was not fully investigated. Hence, the constraints based on the simple models may not represent the true physical conditions of the outflows. Second, the models are primarily used to reproduce the 1D or 2D velocity distribution but not compared with velocity dispersion \citep[e.g.,][]{2013ApJS..209....1F}. Third, statistical constraints are still
missing since the previous studies are limited to a small number of AGNs. To better constrain the NLR kinematics and compare with observed properties, it is of importance to explore the effect of the intrinsic physical parameters, e.g., launching velocity, opening angle, and inclination
of the cone, and to apply the models to a large number of AGNs using full kinematic information (e.g., velocity and velocity dispersion).

\begin{figure*}
\centering
\includegraphics[width=0.95\textwidth]{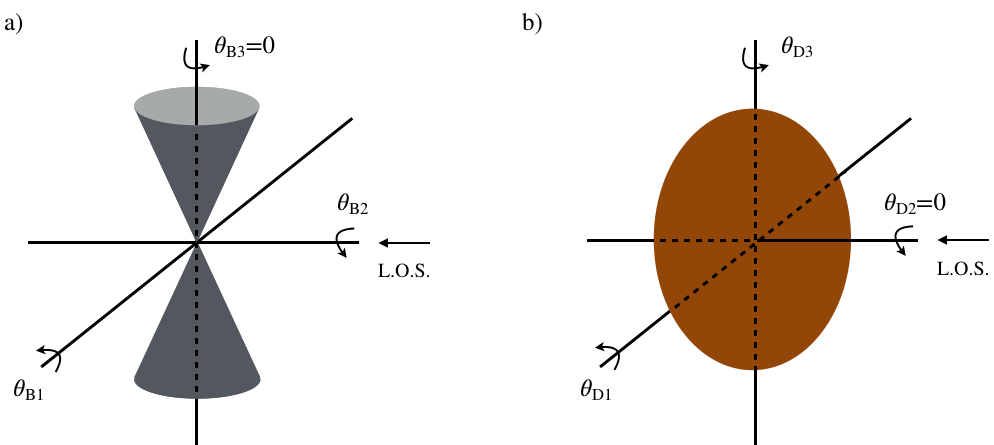}
\caption{Rotational angles of a) the bicone and b) the dust plane in the body-fixed frame. The LOS is along the B2 (D2) axis when there is no rotation in the bicone (dust plane).  Both $\theta_{B3}$ and $\theta_{D2}$ are negligible due to the axisymmetric geometry of the bicone and the dust plane, respectively.}
\label{fig:euler}
\end{figure*}

In the first of a series of papers, we have performed a statistical study to constrain the properties of ionized gas outflows and their relation to
AGN energetics, using a sample of $\sim$39,000 type 2 AGNs selected from Sloan Digital Sky Survey (SDSS) at redshift $z<$0.3 \citep[][hereafter Paper I]{2016ApJ...817..108W}. In particular, the flux-weighted velocity (i.e., first moment) and velocity dispersion (i.e., second moment) of the [\OIII] line were carefully measured, defining the characteristic V-shape distribution in the velocity--velocity dispersion (VVD) diagram,
which reflects the intrinsic properties of biconical outflows as manifested by the [\OIII] line profiles in the spatially integrated fiber spectra \citep[see also][]{2014ApJ...795...30B}. The measured [\OIII] velocities ($v_{\text{[O III]}}$) and velocity dispersions ($\sigma_{\text{[O III]}}$) of this large sample of AGNs can be better interpreted
by comparing with the prediction of the bicone models. 

In this paper, we present biconical outflow models combined with a thin dust plane in 3D, by adopting the model from \citet{2010ApJ...708..419C}. We explore the effect of various physical properties of the bicone and the dust plane on the gas kinematics, and then reproduce
the observed kinematics of  [\OIII] using the large sample of $\sim$39,000 type 2 AGNs from Paper I. By comparing the observed velocity
and velocity dispersion of [\OIII] with the simulated values, we constrain the intrinsic properties of the biconical outflows in type 2 AGNs.

The paper is constructed as follows. In Section \ref{sec:modeling}, we describe the 3D biconical outflow models. In Section \ref{sec:test}, we test the effect of each model parameter on the integrated velocity and velocity dispersion. In Section \ref{sec:profile}, we simulate the emission-line profiles using different parameters. In Section \ref{sec:grid}, we construct model grids to compare with the observed VVD diagram of type 2 AGNs. Section \ref{sec:mc} presents the results of Monte Carlo simulations on the VVD distribution. Section \ref{sec:discussion} provides our discussion and Section \ref{sec:summary} summarizes our findings. In this paper, we adopt a standard $\Lambda$CDM cosmology, i.e., H$_0$ = 70 \kms\ Mpc$^{-1}$, $\Omega_{\Lambda}$ = 0.73, and $\Omega_{\text{m}}$ = 0.27.

\section{Biconical outflow models}
\label{sec:modeling}
We construct a bicone model with two axisymmetric cones, whose apex is located at the nucleus as presented in Figure \ref{fig:bicone}. The cones extend with an outer half-opening angle ($\theta_{\text{out}}$) and have an inner hollow region, defined by an inner half-opening angle ($\theta_{\text{in}}$). We add a thin dust plane representing the dusty stellar disk in the host galaxy, as suggested by the high-resolution imaging and spectroscopic observations \citep[e.g.,][]{2010ApJ...708..419C,2013ApJS..209....1F,2016ApJ...819..148K}, which obscure a part of the cones. For simplicity, we assume that the dust plane has wavelength-independent opacity, since we consider the dust effect on a single emission line, e.g., [\OIII] in a narrow spectral range.  
We define the extinction level of the dust plane with $A$ (i.e., 0 \% $\le A \le$ 100 \%). 
For simplicity, we set the length of the cone $D$ to unity in arbitrary units and the length of the dust plane as a factor of two larger than the length of the cone. 

We assume that the orientations of the bicone and the dust plane are independent of each other, as suggested from observations of local AGNs \citep[e.g.,][]{2005ARA&A..43..769V,2015ApJ...806...84L}. In 3D, we can describe the random orientation of a bicone with three rotational angles, i.e., $\theta_{B1}$, $\theta_{B2}$, and $\theta_{B3}$, in the body-fixed frame (see Figure \ref{fig:euler}). However, only two rotational angles are needed (i.e., $\theta_{B1}$ and $\theta_{B2}$), since the bicone geometry is axisymmetric with respect to  $\theta_{B3}$.
By combining the random rotation of $\theta_{B1}$ and $\theta_{B2}$, we can define the bicone inclination angle ($i_{\text{bicone}}$) from the plane of the sky and the position angle (PA$_{\text{bicone}}$) in the plane of the sky as follows.

\begin{equation}
\begin{split}
i_{\text{bicone}} = S \cos^{-1} {\sqrt{ \sin^2 \theta_{B2} + (\cos \theta_{B1} \cos \theta_{B2}})^2 } \\ ~~S = 
\begin{cases}
+1 & \text{if~ $\sin \theta_{B1} \cos \theta_{B1} > 0$} \\
-1 & \text{if~ $\sin \theta_{B1} \cos \theta_{B1} < 0$}, 
\end{cases}  
\end{split}
\end{equation}

\begin{equation}
\text{PA}_{\text{bicone}} = \tan^{-1} [\sin \theta_{B2} / (\cos \theta_{B1} \cos \theta_{B2}) ],
\end{equation}
where $i_{\text{bicone}}$ and PA$_{\text{bicone}}$ range from $-$90\degree\ to +90\degree. 
The bicone inclination angle and PA are the key parameters to characterize the LOS velocity
distribution observed by an observer (see Figure \ref{fig:bicone}). Note that $\theta_{B1}=i_{\text{bicone}}$ if $\theta_{B2}=0$\degree, while $\theta_{B2}$=PA$_{\text{bicone}}$ if $\theta_{B1}=0$\degree. We define PA$_{\text{bicone}}=0$\degree\ if the bicone axis is oriented north-south.  
In this study, we focus on the range of $i_{\text{bicone}}$ from $-$40\degree\ to $+$40\degree\ as a probable range for type 2 AGNs \citep{2014MNRAS.441..551M} following the expectations from the AGN unification model \citep[e.g.,][]{1995PASP..107..803U}.   

Similarly, we use two different rotational angles for the dust plane (i.e., $\theta_{D1}$ and $\theta_{D3}$) to arbitrarily rotate the dust plane while  the dust plane is axisymmetric with respect to  $\theta_{D2}$ (see Figure \ref{fig:euler}).  
By combining $\theta_{D1}$ and $\theta_{D3}$, we define the inclination of the dust plane ($i_{\text{dust}}$) from the plane of the sky and the position angle (PA$_{\text{dust}}$) in the plane of the sky as follows.

\begin{equation}
\begin{split}
i_{\text{dust}} = S \cos^{-1} [\cos \theta_{D1} \cos \theta_{D3}] \\ ~~S = 
\begin{cases}
+1 & \text{if~ $\sin \theta_{D1} \cos \theta_{D1} > 0$} \\
-1 & \text{if~ $\sin \theta_{D1} \cos \theta_{D1} < 0$}, 
\end{cases}  
\end{split}
\end{equation}

\begin{equation}
\text{PA}_{\text{dust}} = 90^\circ + \tan^{-1} [\sin \theta_{D3} / (\sin \theta_{D1} \cos \theta_{D3}) ],
\end{equation}
where $i_{\text{dust}}$ and PA$_{\text{dust}}$ range from $-$90\degree\ to +90\degree. We note that $\theta_{D1}=i_{\text{dust}}$ if $\theta_{D3}=0$\degree, and PA$_{\text{dust}}=0$\degree\ if the major axis of the dust plane is oriented north-south.  
 
We assume the radial profiles of velocity ($v_{d}$) and flux ($f_{d}$) of the outflows as a function of distance $d$ from the nucleus. For the velocity profile, we assume three simple cases: (1) a linear increase from zero to a maximum velocity $v_\text{{max}}$ ($v_{d}=kd$, where $v_{d}=v_{\text{max}}$ at $d$=$D$=1; (2) a linear decrease ($v_{d}=v_{\text{max}}-kd$, where $v$=0 at $d$=$D$=1; and (3) a constant velocity ($v_{d}=v_{\text{max}}$). In the case of the flux, we assume that the flux exponentially decreases as a function of distance from the center as $f_{d} =f_{n} e^{-\tau (d/D)}$, where $f_{n}$ is an arbitrary flux value at the nucleus, and $\tau$ sets the shape of the flux profile. For example, if $\tau$=5, then the flux at the end of the cone (i.e., d=D) becomes 1/150 of the central value. 
We use various flux profiles, using a range of $\tau$ values based on the observations of local Seyfert galaxies \citep{2013ApJS..209....1F}, which showed the exponential decrease of the radial flux profiles of (bi-)conical structures. For example, NGC 4051 has about two orders of magnitude smaller flux at the edge of the bicone structure than the flux at the nucleus, which corresponds to $\tau=\sim$5 in our models. For comparison, we also use the case of a constant flux along the cone (i.e., $\tau$=0). Table \ref{tbl-1} presents the summary of the model parameters. 

\begin{table}
\center
\caption{Model parameters of the biconical outflows \label{tbl-1}}
\begin{tabular}{ll}
\tableline
\tableline
Parameter & Description \\
\tableline
PA$_{\text{bicone}}$ & position angle of bicone axis \\
$i_{\text{bicone}}$  & inclination of bicone axis \\
$\theta_{\text{out}}$ & outer half-opening angle \\
$\theta_{\text{in}}$ & inner half-opening angle \\
$f_{d}$ & radial flux profile \\
$v_{d}$ & radial velocity profile \\
$v_{\text{max}}$ & maximum velocity\\
\tableline
PA$_{\text{dust}}$ & position angle of dust plane \\
$i_{\text{dust}}$ & inclination of dust plane \\
$A$ & amount of dust extinction \\
\tableline
\end{tabular}
\end{table}

\begin{figure*}
\centering
\begin{tabular}{cc}
\includegraphics[height=5.4cm]{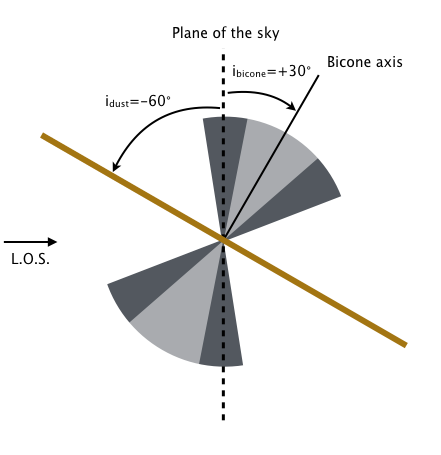}  & \includegraphics[height=5.4cm]{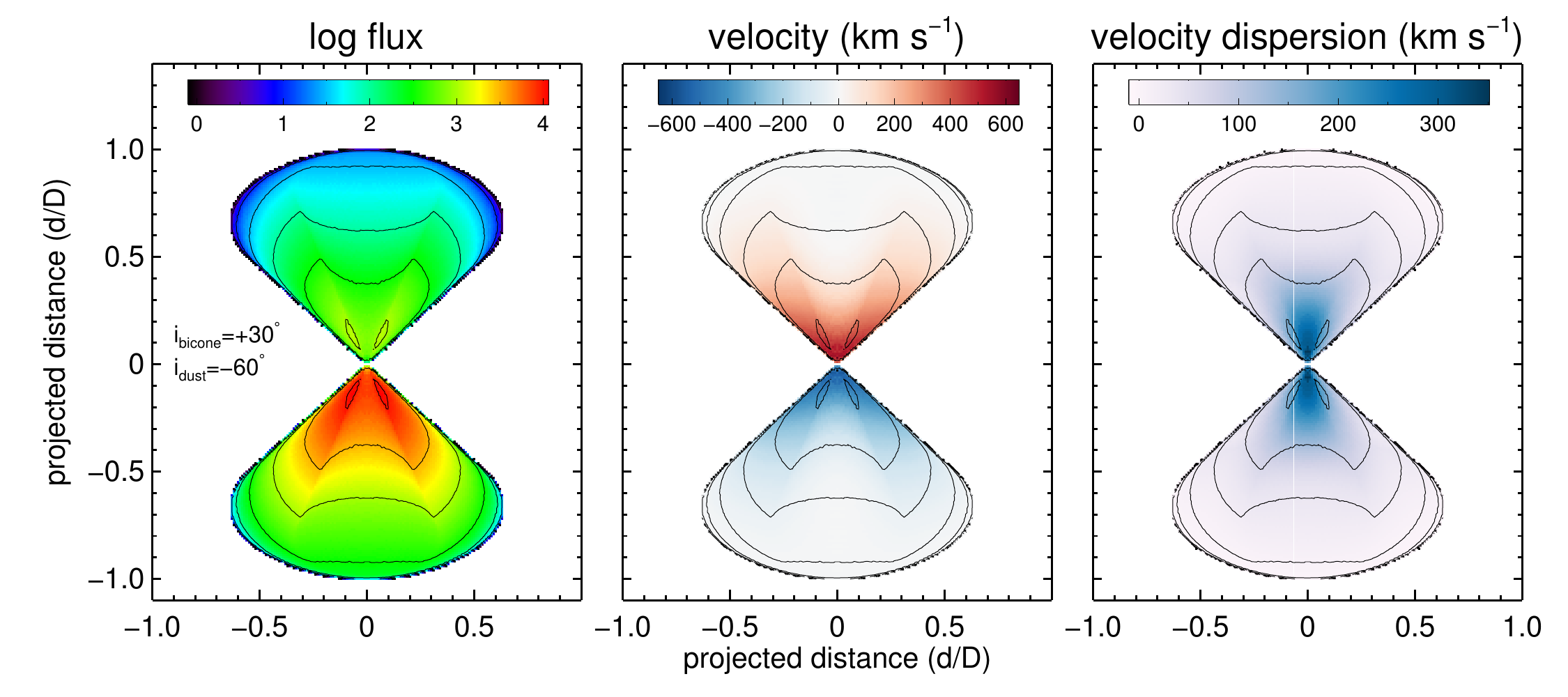} \\
\includegraphics[height=5.4cm]{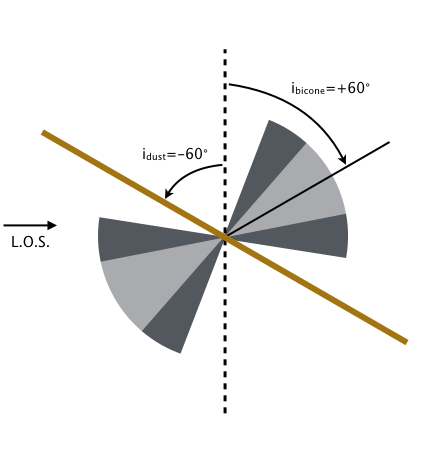}  & \includegraphics[height=5.4cm]{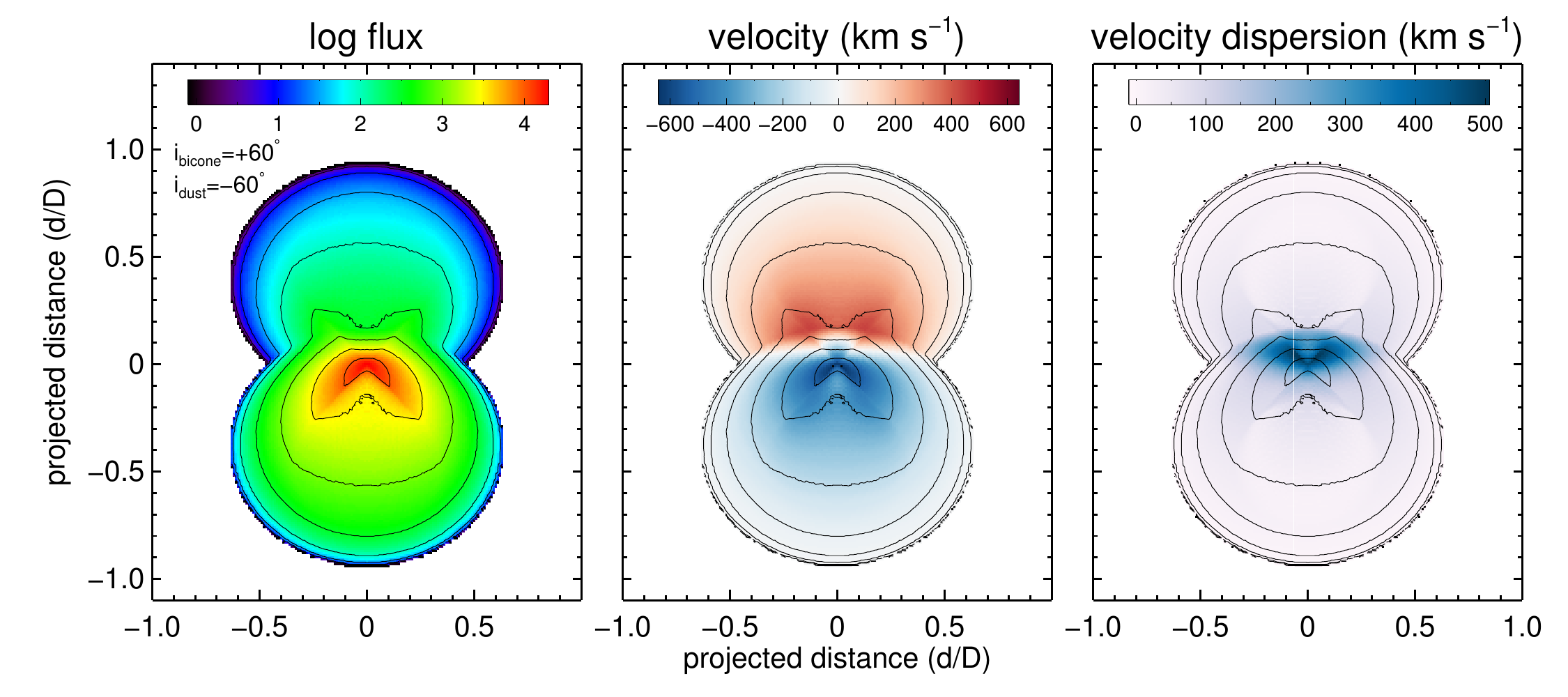} \\
\end{tabular}

\caption{(Left panel) A cartoon for the geometry of the bicone and the dust plane with $i_{\text{dust}}$=$-$60\degree, $\theta_{\text{out}}$=40\degree, $\theta_{\text{in}}$=20\degree, and PA$_{\text{bicone}}$=PA$_{\text{dust}}$=0\degree\ for simplicity, but a different $i_{\text{bicone}}$=+30\degree\ (top) and $i_{\text{bicone}}$=+60\degree\ (bottom). (Right panel) The 2D projected maps of flux (left), flux-weighted velocity (middle), and velocity dispersion (right) based on the bicone model on the left panel. The intrinsic velocity is linearly decreasing from $v_{\text{max}}$=1000 \kms, and the flux is exponentially decreasing with $\tau$=5. Dust extinction is 90\%. Flux maps are in logarithmic scales with arbitrary units. Corresponding flux contours are overlaid on the flux, velocity, and velocity dispersion maps.}
\label{fig:2dmap}
\end{figure*}
  
Since the dust plane hides a part of the bicone, the relative position of the dust plane with respect to the bicone can alter the observed velocity profiles integrated along the LOS. For example, if there is no extinction, the flux-weighted velocity from the outflow would be zero due to the cancelation of the velocities of approaching and receding gas, regardless of the bicone geometry. On the other hand, if one cone is entirely hidden by the dust plane while the other cone is unaffected by dust (for example, see the geometries of the bicone and the dust plane in Figure \ref{fig:bicone}), we observe an approaching or a receding cone depending on the orientation. As a result, the velocity shift integrated over the LOS will increase either negatively or positively, while the velocity dispersion decreases compared to the case of no extinction. 
We will examine the effect of dust extinction in detail in Section \ref{sec:bincl}. 

Based on the orientation of the bicone and the dust plane as described above, 
we construct a 2D flux map $F(x,y)$, which is calculated by integrating the flux along the LOS (z axis) as follows:

\begin{equation}
\begin{split}
F(x,y) = \int f(x,y,z)dz, \\ ~~ f(x,y,z)=
\begin{cases}
f(x,y,z) & \text{if $z< z_{\text{dust}}$} \\
f(x,y,z) \times A & \text{if $z \ge z_{\text{dust}}$}, 
\end{cases}  
\end{split}
\end{equation}
where $x$ and $y$ represent the position in the plane of the sky, while the $z$ axis is defined along the LOS, $f(x,y,z)$ is the flux of a cell in 3D, $z_{\text{dust}}$ is the distance of the dust plane along the LOS from an observer at $z$=0, and $A$ is the amount of dust extinction, ranging $0\% < A < 100\%$. A flux-weighted 2D velocity $V(x,y)$ along the LOS (z axis) can be also calculated as  

\begin{equation}
V(x,y) = \frac{\int v_p(x,y,z)f(x,y,z)dz}{\int f(x,y,z)dz},
\end{equation}
where $v_p(x,y,z)$ is the projected velocity onto the ($x,y$) plane at a given inclination angle of each cell, i.e., $v_p = v_{d}(x,y,z)\times$cos$i'$, where $i'$ ranges from  $i_{\text{bicone}}+\theta_{\text{out}}$ to $i_{\text{bicone}}-\theta_{\text{out}}$ depending on the location within the cone. Finally, a 2D projected velocity dispersion $\sigma (x,y)$ is calculated as follows.

\begin{equation}
\sigma^2 (x,y) = \frac{\int v_p^{2} (x,y,z)f(x,y,z)dz}{\int f(x,y,z)dz} - V^2 (x,y).
\end{equation}
 
In Figure \ref{fig:2dmap}, we present two examples of 2D maps projected from 3D bicone models,
respectively, with two different inclination angles of the bicone: $i_{\text{bicone}}$=+30\degree\ (top) and $i_{\text{bicone}}$=+60\degree\ (bottom).
For simplicity, we use $i_{\text{dust}}$=$-$60\degree, $\theta_{\text{out}}$=40\degree, $\theta_{\text{in}}$=20\degree, and PA$_{\text{bicone}}$=PA$_{\text{dust}}$=0\degree\ (left panels in Fig \ref{fig:2dmap}). We assume that the intrinsic velocity linearly decreases from $v_{\text{max}}$=1000 \kms, while the flux exponentially decreases with $\tau$=5, and dust extinction A= 90\%. Flux maps are expressed with contours of one order of magnitude intervals, which are also overlaid in the velocity and velocity dispersion maps. Note that the receding cone is much fainter than the approaching cone due to dust obscuration. If there is no overlap between the two cones along the LOS due to low $i_{\text{bicone}}$ (top panel), the velocity map is clearly separated into positive and negative values, while velocity dispersion is higher near the center than at the outskirts of the bicone. In contrast, if there is an overlapping region between the two cones along the LOS due to high $i_{\text{bicone}}$ (bottom panel), some part of the overlap region show a zero velocity and a large velocity dispersion in the projected plane. The detailed analysis of the 2D results compared to the spatially resolved measurements based on integral-field spectroscopy \citep[e.g.,][]{2016ApJ...819..148K} will be presented in a forthcoming paper.

The main purpose of this work is to obtain flux-weighted velocity and velocity dispersion integrated over the outflow region for comparing with the observed values of a statistical sample, e.g, SDSS type 2 AGNs (see Paper I).
From the 2D projected flux, velocity, and velocity dispersion maps, we calculate the integrated velocity $v_{\text{int}}$ and velocity dispersion $\sigma_{\text{int}}$ of the outflows, respectively, as follows.

\begin{equation}
v_{\text{int}} = \frac{\iint V(x,y)F(x,y)dxdy}{\iint F(x,y)dxdy},
\end{equation}

\begin{equation}
\sigma^2 _{\text{int}} = \frac{\iint V^{2}(x,y)F(x,y)dxdy}{\iint F(x,y)dxdy} - v^2_{\text{int}}.
\end{equation}
These $v_{\text{int}}$ and $\sigma_{\text{int}}$ can be directly compared to the observed values to constrain the intrinsic properties of outflows.

\section{Tests on the effects of parameters}
\label{sec:test}
In this section, we investigate how the integrated velocity $v_{\text{int}}$ and velocity dispersion $\sigma_{\text{int}}$ vary as a function of the input parameters of the bicone as summarized in Table \ref{tbl-1}. This will help us to understand the underlying physics and intrinsic properties of the outflows. 
Using biconical outflow models on 17 local Seyfert galaxies, \citet{2013ApJS..209....1F} estimated that $v_{\text{max}}$ ranges from 130 \kms\ to 2000 \kms, $\theta_{\text{out}}$ ranges from 25\degree\ to 62\degree, and $\theta_{\text{in}}$ is about 25\% to 95\% of $\theta_{\text{out}}$. For simplicity, we initially fix a couple of parameters motivated from these results, i.e., $\theta_{\text{out}}$=40\degree, $\theta_{\text{in}}$=20\degree, PA$_{\text{bicone}}$, PA$_{\text{dust}}=0$\degree. We assume a constant velocity of 1000 \kms\ and a constant flux as a function of distance. 

\begin{figure}
\centering
\includegraphics[width=0.49\textwidth]{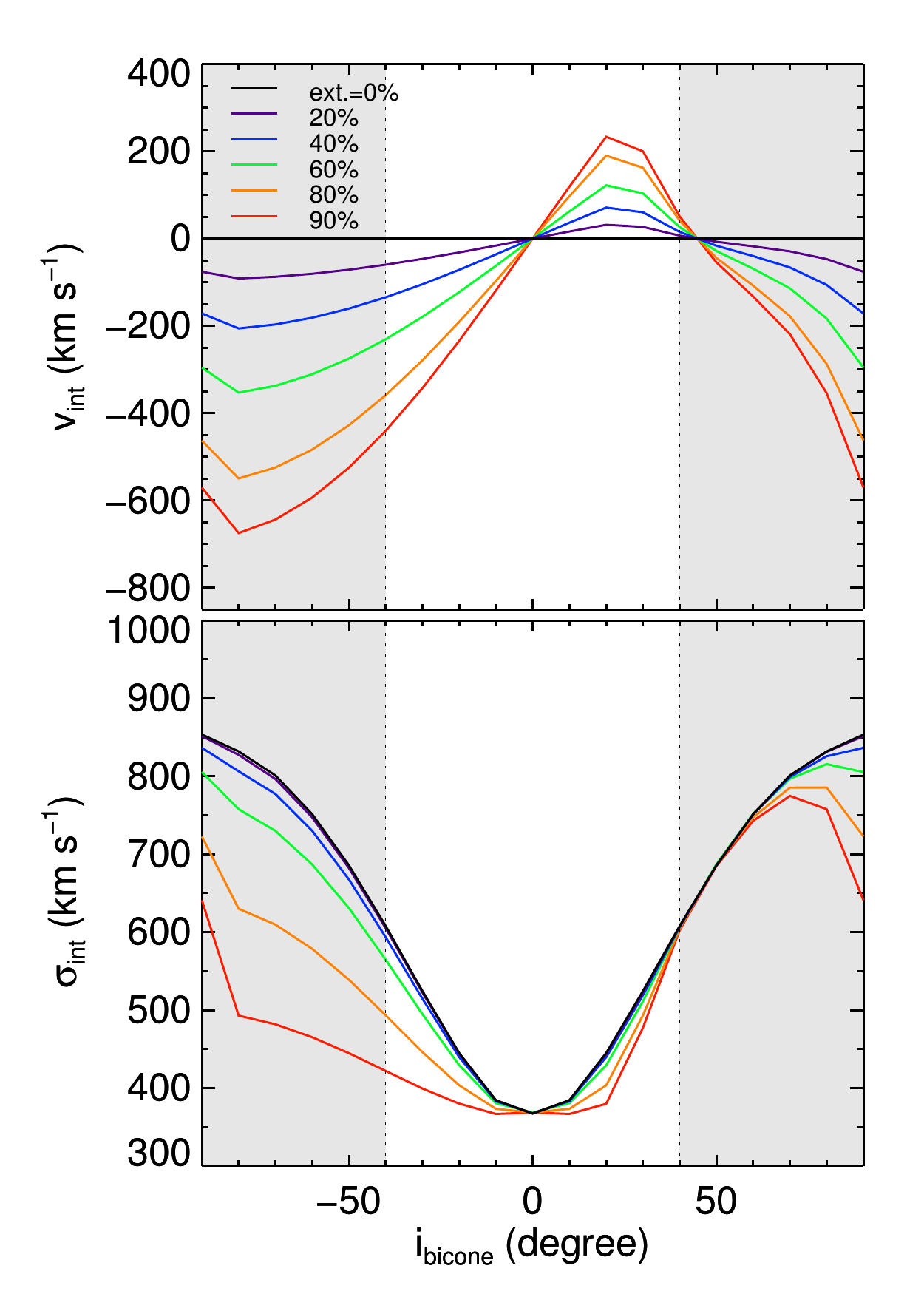}
\caption{Integrated velocity (top) and velocity dispersion (bottom) as a function of bicone inclination with a different amount of dust extinction. The unshaded region represents the probable range of the bicone inclinations for type 2 AGNs ($-$40\degree-- +40\degree). We assume that the intrinsic velocity is constant as 1000 \kms, and the radial flux is also constant, $\theta_{\text{out}}$=40\degree, $\theta_{\text{in}}$=20\degree, and $i_{\text{dust}}$=60\degree.}
\label{fig:1dincl}
\end{figure}

\begin{figure}
\centering
\includegraphics[width=0.49\textwidth]{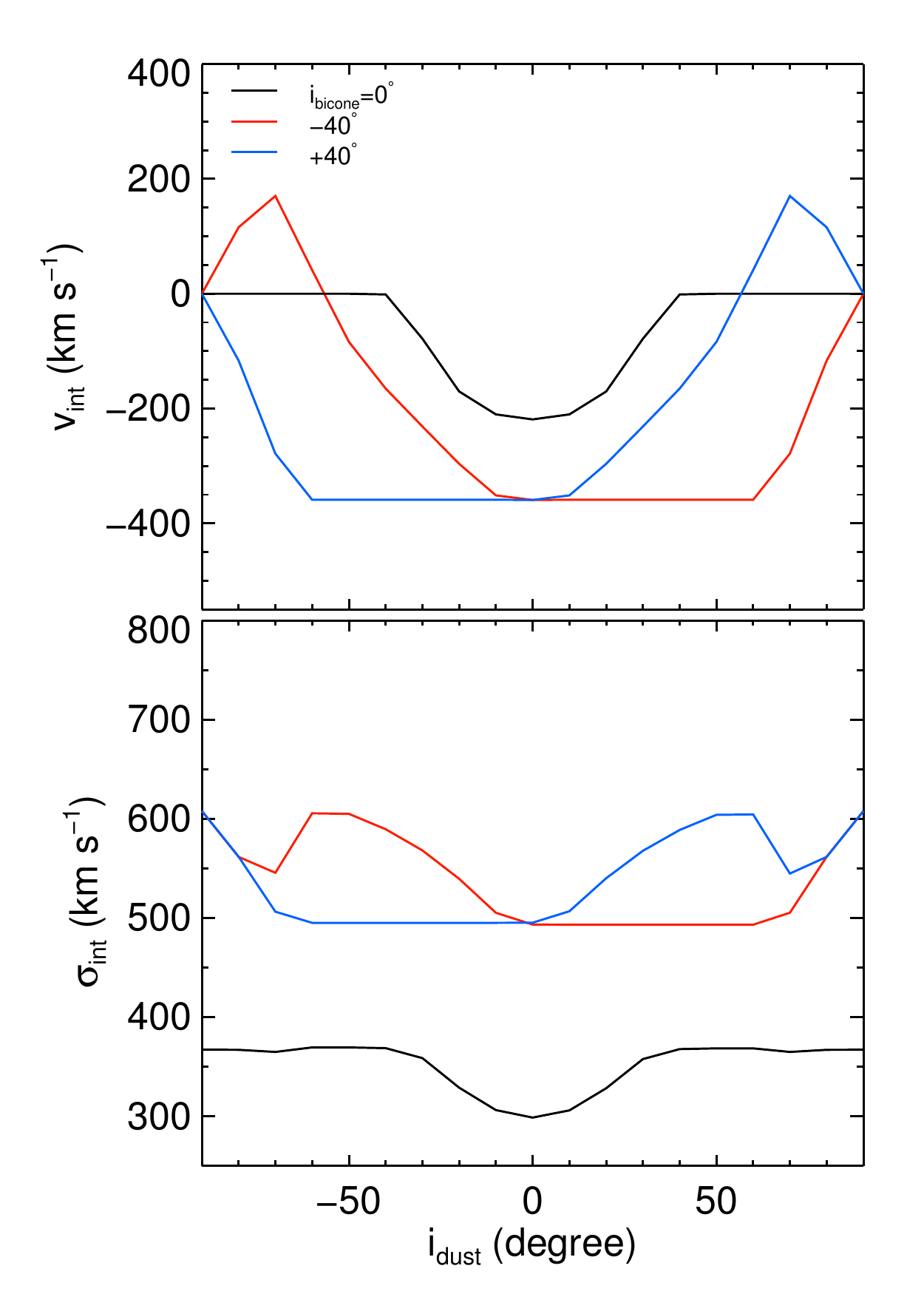}
\caption{Integrated velocity (top) and velocity dispersion (bottom) as a function of dust inclination for three different bicone inclinations, i.e., $-$40\degree\ (red), 0\degree\ (black), and +40\degree\ (blue). We assume that the intrinsic velocity is constant as 1000 \kms, the radial flux is also constant, $\theta_{\text{out}}$=40\degree, $\theta_{\text{in}}$=20\degree, and the dust extinction $A$=50\%.}
\label{fig:dincl}
\end{figure}

\begin{figure}
\centering
\includegraphics[width=0.49\textwidth]{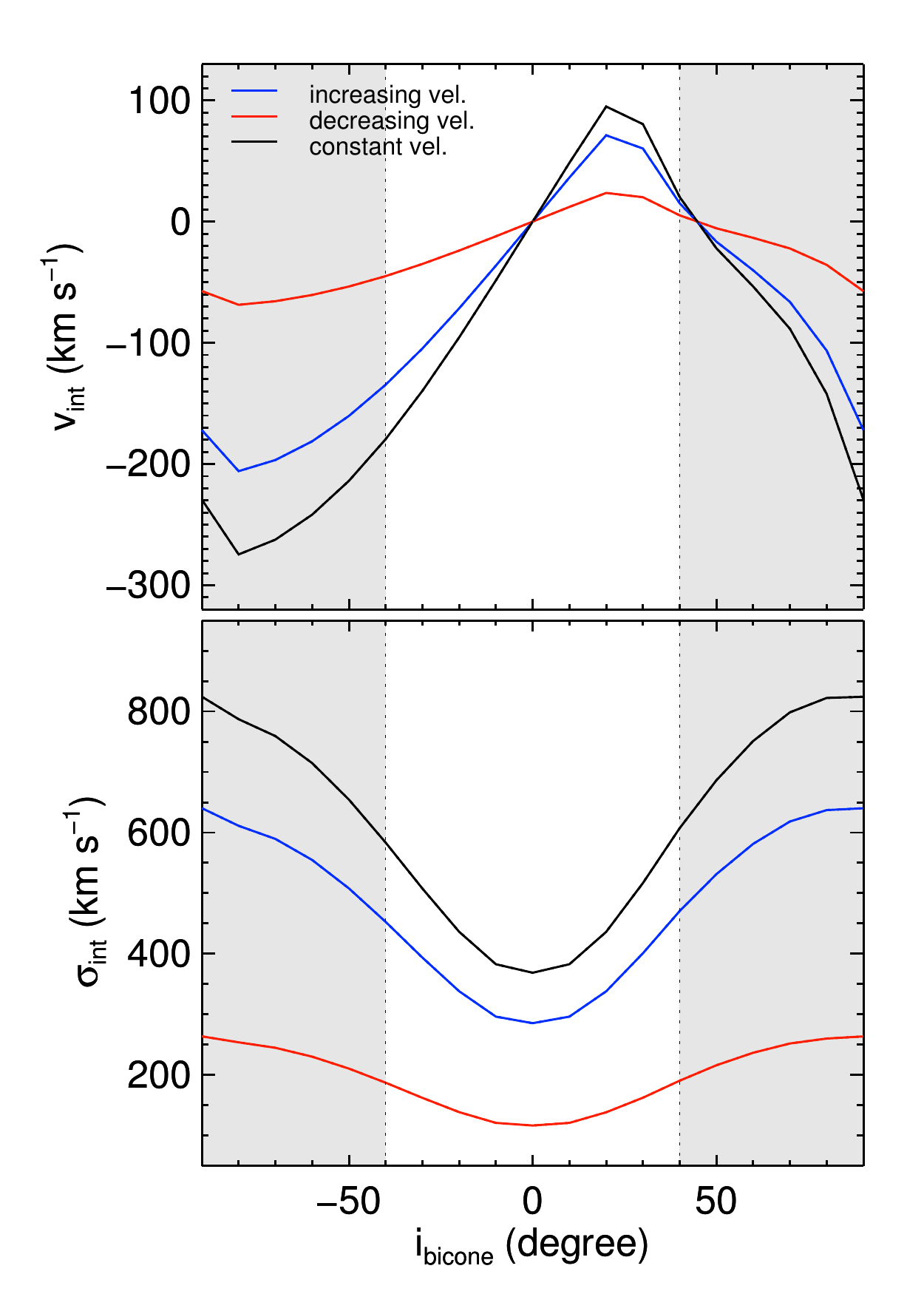}
\caption{Integrated velocity (top) and velocity dispersion (bottom) as a function of bicone inclination with a different velocity structure, i.e., constant velocity (black), linear increasing velocity (blue), and linear decreasing velocity (red). The unshaded region represents a probable range of bicone inclinations for type 2 AGNs ($-$40\degree-- +40\degree). We assume that the radial flux is constant, the maximum velocity is 1000 \kms, $\theta_{\text{out}}$=40\degree, $\theta_{\text{in}}$=20\degree, $i_{\text{dust}}$=+60\degree, and the dust extinction $A$=50\%.}
\label{fig:1dvp}
\end{figure}

\begin{figure}
\centering
\includegraphics[width=0.49\textwidth]{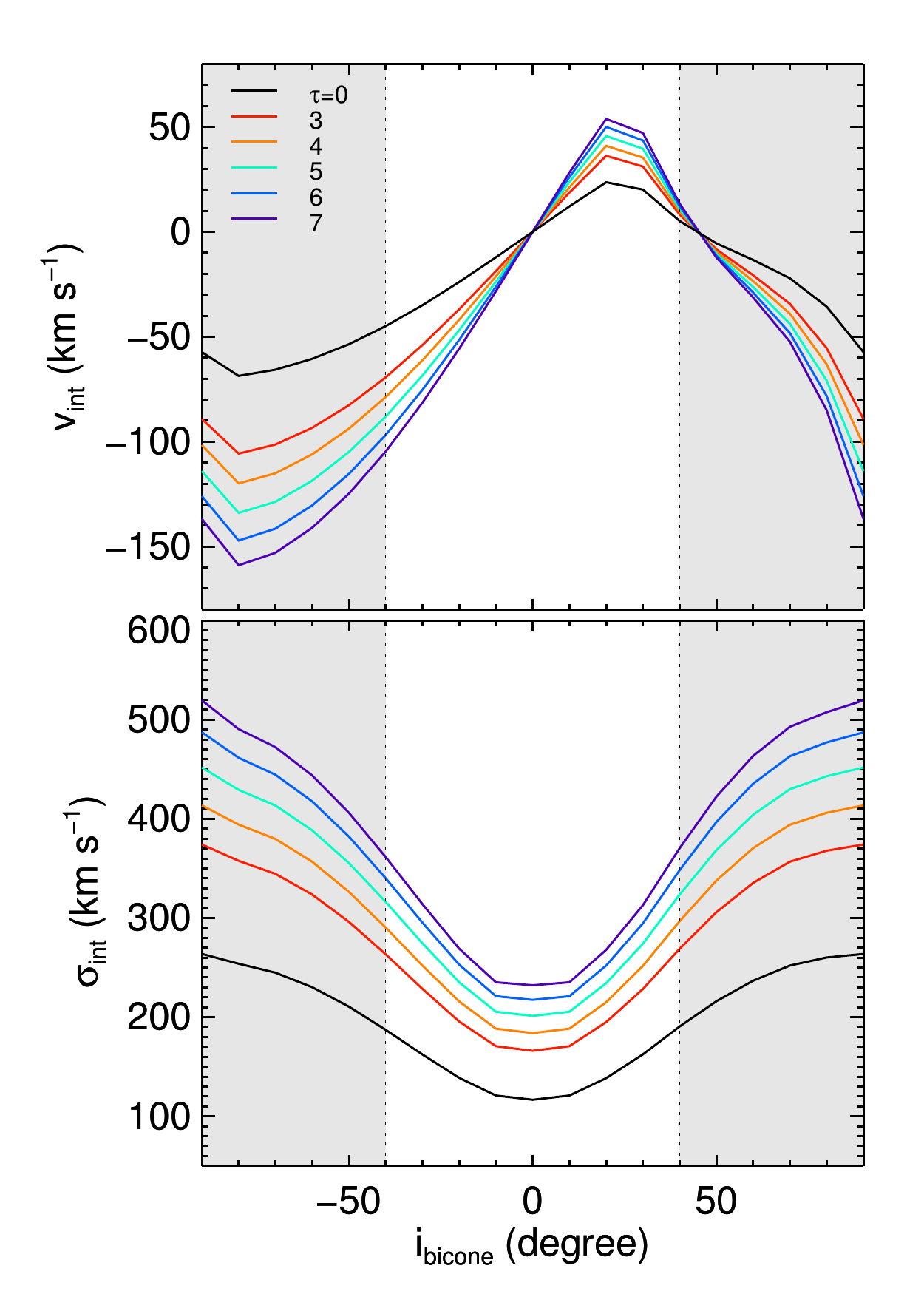}
\caption{Integrated velocity (top) and velocity dispersion (bottom) as a function of bicone inclination with a different $\tau$ in a radial flux distribution function, i.e., $f_{d} = f_{n}e^{-\tau d}$. The unshaded region represents a probable range of bicone inclinations for type 2 AGNs ($-$40\degree-- +40\degree). We assume that the intrinsic velocity is linearly decreasing from a maximum velocity of 1000 \kms, $\theta_{\text{out}}$=40\degree, $\theta_{\text{in}}$=20\degree, $i_{\text{dust}}$=+60\degree, and the dust extinction $A$=50\%.}
\label{fig:1dfp}
\end{figure}

\subsection{Bicone inclination and dust extinction}
\label{sec:bincl}
For a given intrinsic velocity $v_{\text{max}}$, two major parameters affect the integrated gas kinematics: (1) the bicone inclination $i_{\text{bicone}}$; 
and (2) the amount of dust extinction $A$.
Here we examine the effect of the two parameters on the integrated kinematics.  

First of all, we consider a simple case of no dust extinction. Since the velocities of the approaching and receding gas cones cancel each other out, the integrated velocity $v_{\text{int}}$ remains zero regardless of $i_{\text{bicone}}$  (black line on the top panel in Figure \ref{fig:1dincl}). On the other hand, the velocity dispersion $\sigma_{\text{int}}$ varies as $i_{\text{bicone}}$ changes (black line on the bottom panel), because the projection of approaching and receding gas varies. 
$\sigma_{\text{int}}$ has the smallest value when the bicone axis is parallel to the plane of the sky (i.e., $i_{\text{bicone}}$=0\degree), since the projection to the LOS is highest, hence the LOS velocity distribution (LOSVD) is narrowest. $\sigma_{\text{int}}$ becomes larger as $i_{\text{bicone}}$ becomes higher until the bicone axis becomes parallel to the LOS (i.e., $i_{\text{bicone}}=\pm$90\degree). $\sigma_{\text{int}}$ can be a factor of $\sim$1.5 larger at $i_{\text{bicone}}$=$\pm$40\degree\ (for type 2 AGNs), and a factor of $\sim$2 larger at $i_{\text{bicone}}$=$\pm$90\degree\ (for type 1 AGNs), compared to $\sigma_{\text{int}}$ at $i_{\text{bicone}}$=0\degree.  

Second, we examine the effect of dust extinction as a function of $i_{\text{bicone}}$,
by adopting a fixed inclination of the dust plane $i_{\text{dust}}$=+60\degree.
If there is dust extinction, the flux-weighted velocity $v_{\text{int}}$ does not remain zero due to the asymmetric velocity
distribution along the LOS. Instead, the centroid of the velocity distribution is shifted to either negative or positive values
depending on $i_{\text{bicone}}$ (see the colored lines in the top panel of Figure \ref{fig:1dincl}).
The more dust extinction there is, the larger the $|v_{\text{int}}|$ since dust obscuration makes stronger asymmetric velocity
distribution along the LOS.

For the expected range of $i_{\text{bicone}}$ of type 2 AGNs, i.e., from $-$40\degree\ to +40\degree\ (unshaded area in Figure \ref{fig:1dincl}), $v_{\text{int}}$ becomes more negative as $i_{\text{bicone}}$ becomes inclined from 0\degree\ to $-$40\degree\ toward an observer. It is because the projected velocities increase as the cone becomes more inclined, while the dust obscures the receding cone. Along the other direction of $i_{\text{bicone}}$, on the other hand, $v_{\text{int}}$ becomes larger up to $i_{\text{bicone}}$=+20\degree, and then decreases for higher $i_{\text{bicone}}$, while $v_{\text{int}}$ remains positive due to a partial obscuration up to $i_{\text{bicone}}=\sim$40\degree. When $i_{\text{bicone}}$=+20\degree, the cone-walls are attached to the dust plane. As a result, $v_{\text{int}}$ has the largest positive value due to the full obscuration of the approaching cone. If the bicone is further inclined from +20\degree\ to +40\degree, then both cones are partially obscured by the dust plane, resulting in $v_{\text{int}}$ becoming smaller than the maximum value. If we further incline the bicone from +40\degree, $|v_{\text{int}}|$ becomes larger as the receding cone is more obscured. 

In the case of $\sigma_{\text{int}}$, $\sigma_{\text{int}}$ becomes smaller as the extinction increases at a fixed $i_{\text{bicone}}$ due to a larger differences in flux between the two cones. However, we find that the increasing trend of $\sigma_{\text{int}}$ with $i_{\text{bicone}}$ is, in general, similar to the case without dust extinction, if extinction is moderate, i.e., $A<\sim$60\% (bottom panel in Figure \ref{fig:1dincl}). On the other hand, if the dust extinction is very high, e.g., $A$=90\%, $\sigma_{\text{int}}$ becomes significantly lower than that of the no-extinction case, e.g., $\sim$30\% lower $\sigma_{\text{int}}$ at $i_{\text{bicone}}=-40$\degree\ (type 2 AGNs) and $\sim$40\% lower $\sigma_{\text{int}}$ at $i_{\text{bicone}}=-$80\degree\ (type 1 AGNs). 

\begin{figure}
\centering
\includegraphics[width=0.49\textwidth]{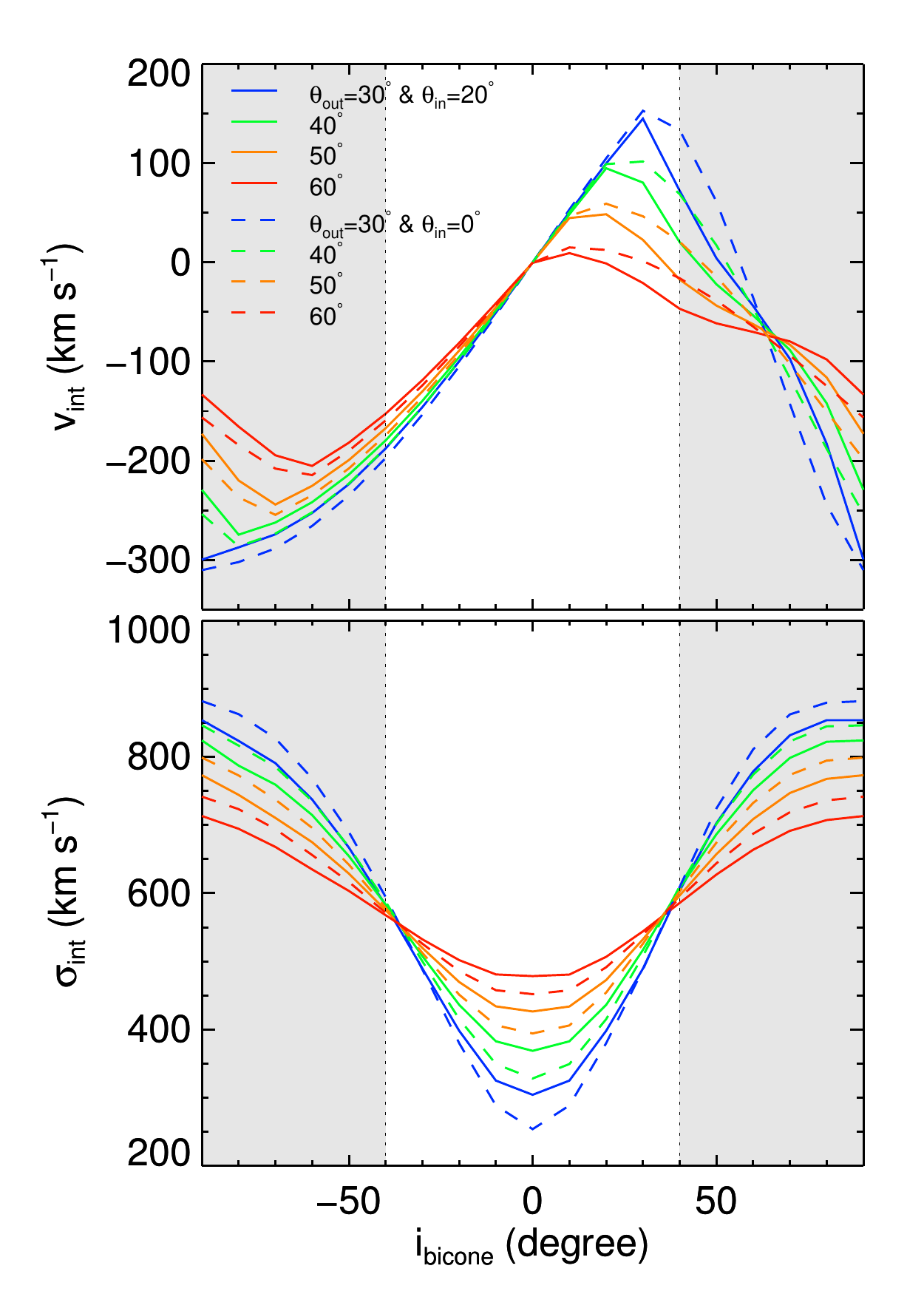}
\caption{Integrated velocity (top panel) and velocity dispersion (bottom panel) as a function of bicone inclination with a different outer opening angle ($\theta_{\text{out}}$) with a fixed inner opening angle ($\theta_{\text{in}}$) of 20\degree\ (solid lines) and 0\degree\ (dashed lines). The unshaded region represents a probable range of bicone inclinations for type 2 AGNs ($-$40\degree-- +40\degree). We assume that the flux is uniform, the velocity is constant with 1000 \kms, $i_{\text{dust}}$=+60\degree, and the dust extinction $A$=50\%.}
\label{fig:1doangle}
\end{figure}

\begin{figure*}
\centering
\includegraphics[width=1.0\textwidth]{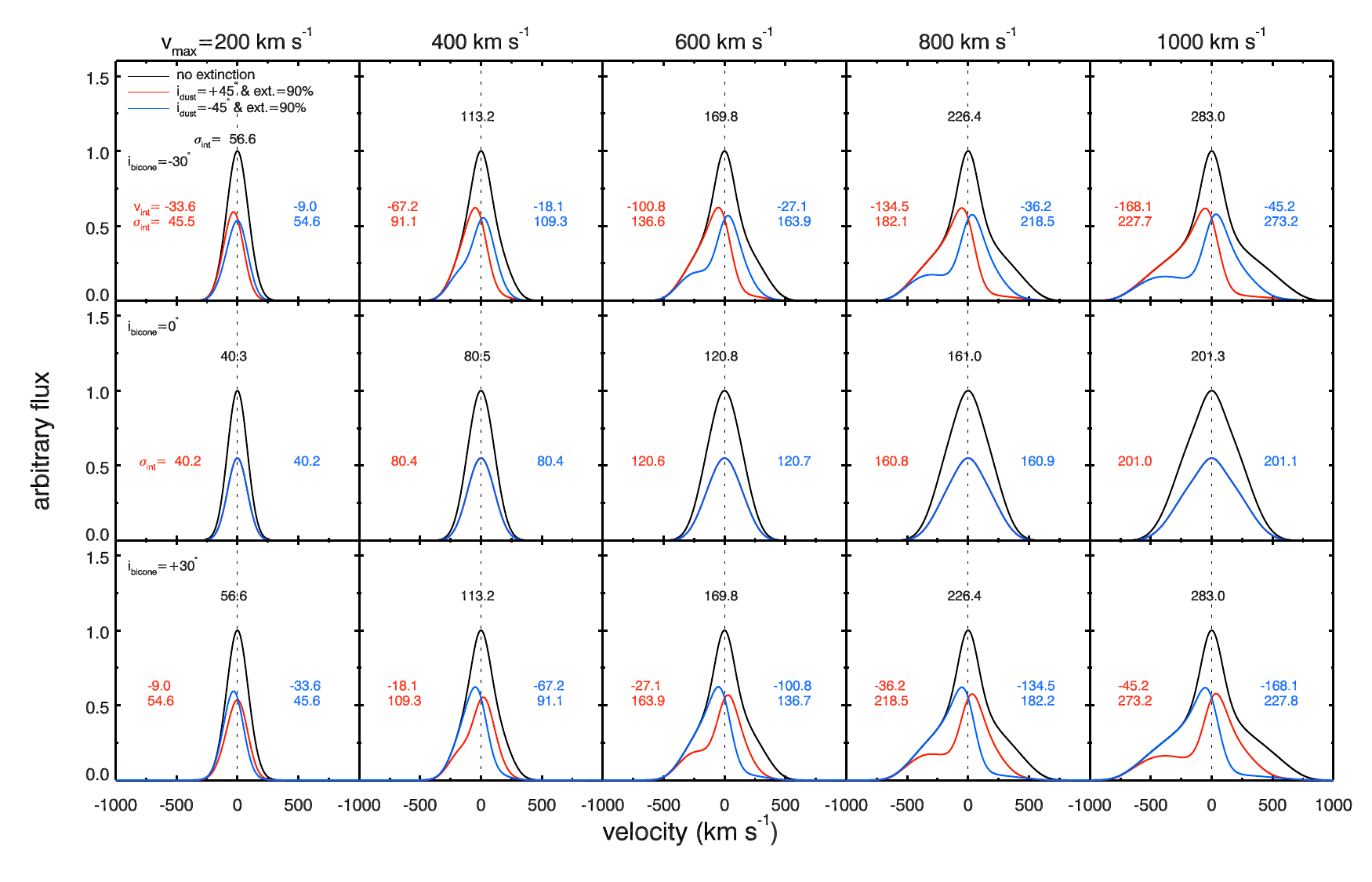}
\caption{Simulated emission-line profiles for different sets of model parameters. We assume the linearly decreasing velocity from $v_{\text{max}}$, and the exponentially decreasing flux profile with $\tau$=5, $\theta_{\text{out}}$=40\degree, and $\theta_{\text{in}}$=20\degree, for simplicity. From left to right, we apply a different $v_{\text{max}}$ from 200 \kms\ to 1000 \kms, respectively. For each line profile, we overlay the calculated $v_{\text{int}}$ and $\sigma_{\text{int}}$. For the case of no dust extinction (black), we omit $v_{\text{int}}$ since $v_{\text{int}}$=0 \kms\ in this case.} 
\label{fig:profile}
\end{figure*}

\begin{figure}
\centering
\includegraphics[width=0.49\textwidth]{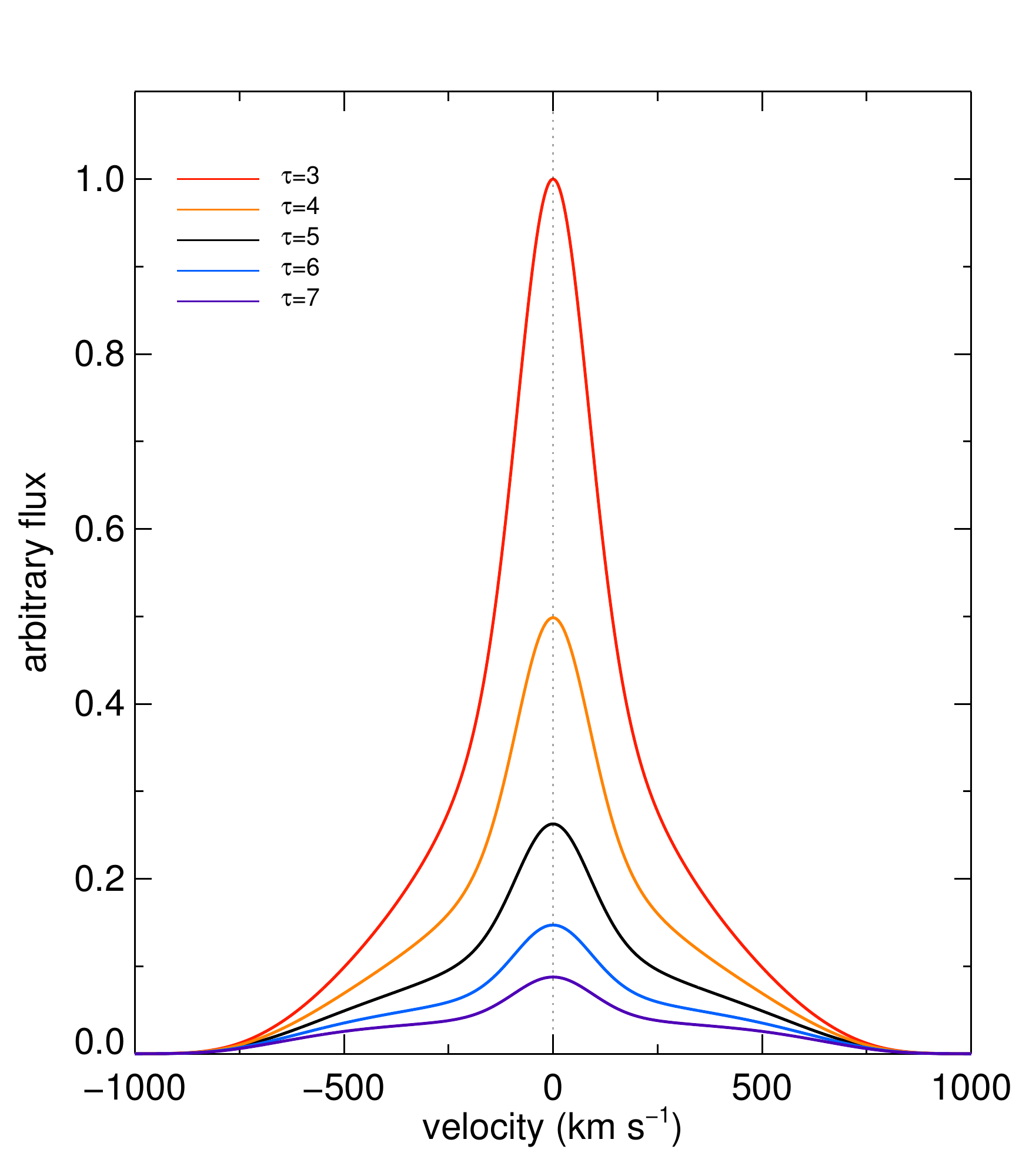}
\caption{Simulated emission-line profiles for different flux gradients, i.e., different values of $\tau$. We assume the linearly decreasing velocity profile with $v_{\text{max}}$=1000 \kms. The bicone inclination is $-$30\degree, $\theta_{\text{out}}$=40\degree, and $\theta_{\text{in}}$=20\degree. The line profile from top to bottom is based on the assumption of $\tau$=[3, 4, 5, 6, 7]}, respectively.
\label{fig:tau}
\end{figure}

\begin{figure}
\centering
\includegraphics[width=0.49\textwidth]{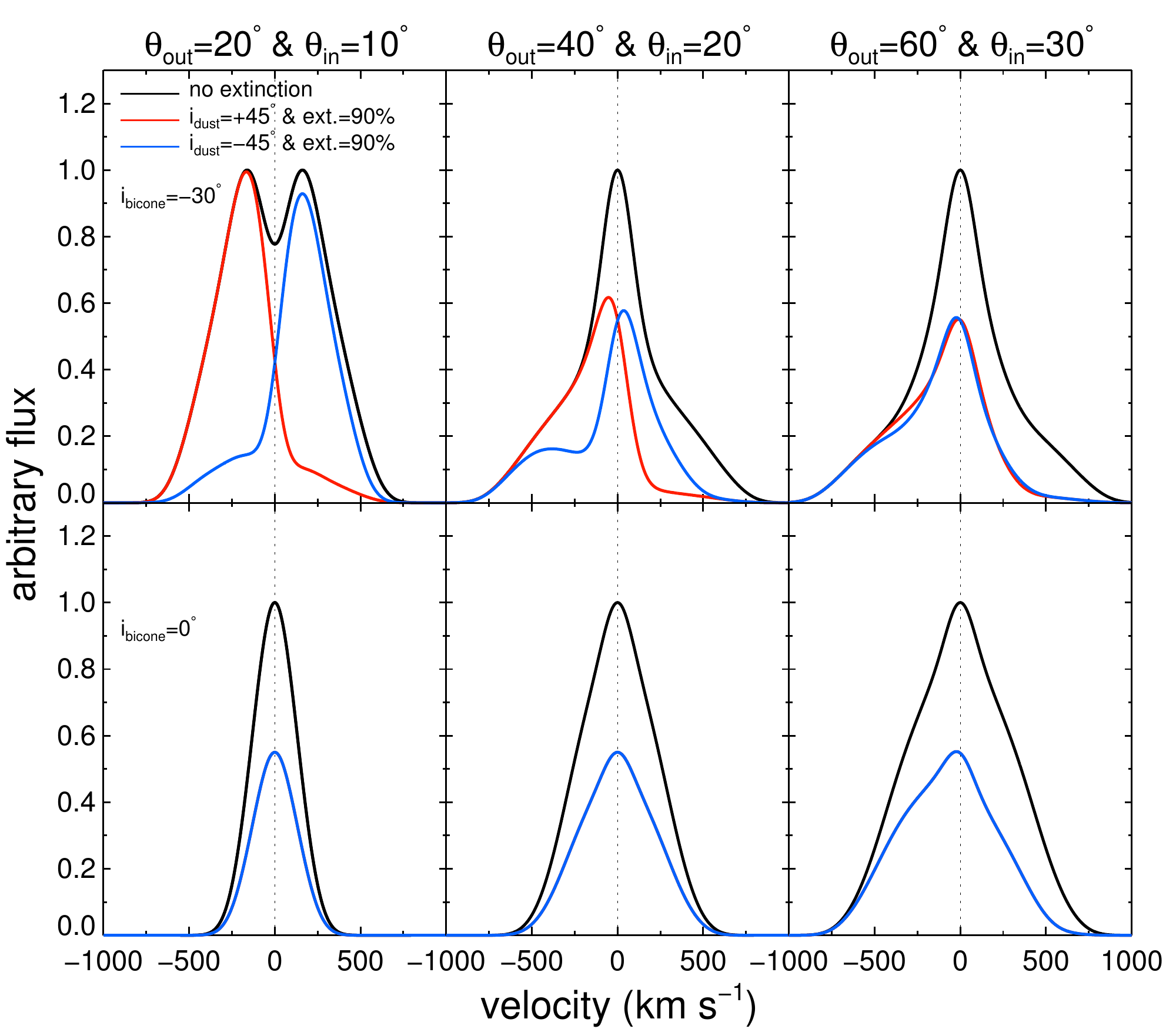}
\caption{Simulated emission-line profiles for different opening angles of $\theta_{\text{out}}$ and $\theta
_{\text{in}}$. We assume $v_{\text{max}}$=1000 \kms, and $\tau$=5. }
\label{fig:oa}
\end{figure}

\subsection{Dust inclination}
Here we test the effect of dust inclination $i_{\text{dust}}$ on both $v_{\text{int}}$ and $\sigma_{\text{int}}$. We use three different $i_{\text{bicone}}$, i.e., $-$40\degree, 0\degree, and +40\degree\ (red, black, and blue lines in Figure \ref{fig:dincl}, respectively) with a fixed dust extinction ($A$=50\%).  

First, we consider $v_{\text{int}}$ for a simple case with $i_{\text{bicone}}$=0\degree\ (top panel). If $i_{\text{dust}}$=0\degree, i.e., the bicone and the dust plane are parallel to the plane of the sky, a half of the bicone is obscured by dust. As a result, $v_{\text{int}}$ has the largest negative value since we mainly see the approaching side of the bicone. $|v_{\text{int}}|$ becomes smaller as the dust plane inclines, and then it becomes zero at $i_{\text{dust}}<-$40\degree\ or >+40\degree\ due to the cancellation of the velocities of approaching and receding gas. 

The trend is more complicated if the bicone is also inclined. If $i_{\text{bicone}}$=$-$40\degree\ and $i_{\text{dust}}$=0\degree, for example, the dust plane obscures the entire receding cone. So the velocity becomes the largest negative value, which is more negative than the case without bicone inclination due to a larger projected velocity. $v_{\text{int}}$ remains the lowest value as the dust plane inclines up to +60\degree, but $|v_{\text{int}}|$ becomes smaller when the dust plane is further inclined $>$+60\degree. The reason for this is that since we limited the size of the dust plane, i.e., twice larger than the bicone, the inclined dust plane cannot obscure the whole cone if $i_{\text{dust}}>\pm$60\degree. Note that this is an artificial effect depending on the assumption of the size of the dust plane. As we further incline the dust plane, it touches the cone-walls at $i_{\text{dust}}=$+110\degree\ (=$-$20\degree). In this geometry, a part of the approaching cone is obscured while the receding cone is fully visible, resulting in $v_{\text{int}}$ becoming the largest positive value. $v_{\text{int}}$ becomes negative again as the dust plane inclines closer to the plane of the sky. We obtain exactly the same results but with an opposite trend if we assume $i_{\text{bicone}}$=+40\degree.

Second, we examine $\sigma_{\text{int}}$ dependence on $i_{\text{dust}}$ (bottom panel). We find that the trend in $\sigma_{\text{int}}$ is similar to the trend in $v_{\text{int}}$ to some extent. Let us consider a simple case with no bicone inclination first. In this case, $\sigma_{\text{int}}$ is the lowest when the dust plane is not inclined, because the LOSVD is not broad due to the obscuration of a half of the bicone. $\sigma_{\text{int}}$ increases as the dust plane is inclined to $\pm$40\degree. If $i_{\text{dust}}$ is further inclined from $\pm$40\degree, then $\sigma_{\text{int}}$ has the largest value since the dust plane is placed outside of the cone.

If $i_{\text{bicone}}$=$-$40\degree\ and $i_{\text{dust}}$=0\degree, $v_{\text{int}}$ along each LOS in the plane of the sky has a larger value compared to the case with $i_{\text{bicone}}=0$\degree, resulting in a factor of approximately two larger $\sigma_{\text{int}}$. As the dust plane inclines, $\sigma_{\text{int}}$ remains flat until $i_{\text{dust}}$=+60\degree. Since the inclined dust plane cannot obscure the entire bicone at $i_{\text{bicone}}>$+60\degree, $\sigma_{\text{int}}$ increases about 20\% as the LOSVD broadens. $\sigma_{\text{int}}$ is the largest at $i_{\text{dust}}$=$\pm$90\degree due to the no-extinction effect. In short, $\sigma_{\text{int}}$ changes in a complex way as $i_{\text{dust}}$ varies, but the values remain, in general, within $\sim$10\% of the mean value.

\subsection{Radial velocity profile}
Here we test three different cases of the intrinsic velocity profiles: (1) a linear increase; (2) a linear decrease; and (3) a constant velocity (Figure \ref{fig:1dvp}). We choose $v_{\text{max}}=1000$ \kms, $i_{\text{dust}}$=+60\degree, and $A$=50\%, for simplicity.

Because we adopt a constant flux as a function of distance, $v_{d}$ in the outer cells is more weighted than that in the inner cells in calculating $v_{\text{int}}$ and $\sigma_{\text{int}}$ since the number of cells is much larger in the outer region than in the inner region in the bicone geometry. If $v_{d}$ increases as a function of distance, it affects $v_{\text{int}}$ and $\sigma_{\text{int}}$ more, compared to the case of decreasing velocity. If $v_{d}$ is not constant, the observed |$v_{\text{int}}$| will be smaller than the case of constant velocity. 

In Figure \ref{fig:1dvp}, the case of decreasing velocity (red lines) shows the smallest $|v_{\text{int}}|$. Conversely, the case of increasing velocity (blue lines) shows a smaller but comparable $|v_{\text{int}}|$ to the case of constant velocity (black lines) over the whole range of $i_{\text{bicone}}$ (top panel). We also find a similar trend in $\sigma_{\text{int}}$ (bottom panel): the case of decreasing velocity has the smallest $\sigma_{\text{int}}$, while the case of increasing velocity has the intermediate $\sigma_{\text{int}}$.

\subsection{Radial flux profile}
Here we consider the effect of the radial flux profile on both $v_{\text{int}}$ and $\sigma_{\text{int}}$. The radial flux profile should be considered together with the velocity profile because the flux of each cell determines the weight for the kinematics calculation (Equations (8) and (9)). We examine two cases of the radial flux profile $f_{d} =f_{n} e^{-\tau d}$: (1) a uniform ($\tau=0$); and (2) an exponential decrease ($\tau>$0). For the latter case, we use five different $\tau$=[3, 4, 5, 6, 7]. We assume $i_{\text{dust}}$=+60\degree, and $A$=50\%, for simplicity. We decrease the radial velocity from $v_{\text{max}}$=1000 \kms\ to see the effect of different $\tau$. If we use a constant radial velocity, we find that $\tau$ does not have any effect on both $v_{\text{int}}$ and $\sigma_{\text{int}}$.  

We expect a steep decline of radial flux for large values of $\tau$. This means that the contribution from the outer region of the bicone, where the velocity is close to zero in this case, decreases with increasing $\tau$. In the meantime, the contribution from the central region, where the velocity is the largest, dominates at larger $\tau$. As a result, larger $\tau$ affects the kinematics more. As shown in Figure \ref{fig:1dfp}, $|v_{\text{int}}|$ is the largest when $\tau$ is the largest (=7), while the $|v_{\text{int}}|$ is the smallest when the flux is uniform (i.e., $\tau$=0). Similarly, $\sigma_{\text{int}}$ is the largest with $\tau$=7 and is the smallest with $\tau$=0 for all $i_{\text{bicone}}$.

%The results demonstrate that the velocity and flux profiles should be considered together in the biconical outflow model.

\subsection{Bicone opening angle}
\label{sec:boa}
Now we investigate the effect of the opening angles on both $v_{\text{int}}$ and $\sigma_{\text{int}}$. Here we assume $i_{\text{dust}}$=+60\degree, and $A$=50\%, for simplicity. We choose the range of $\theta_{\text{out}}$ from 30\degree\ to 60\degree.

First, we fix the size of the hollow region ($\theta_{\text{in}}$=20\degree) inside of the bicone and test the effect of $\theta_{\text{out}}$ (solid lines in Figure \ref{fig:1doangle}). In general, the trend of $v_{\text{int}}$ and $\sigma_{\text{int}}$ as a function of $i_{\text{bicone}}$ is similar to what we see in Section \ref{sec:bincl}. If $\theta_{\text{out}}$ becomes smaller, the maximum positive $v_{\text{int}}$ becomes larger and presents at higher  $i_{\text{bicone}}$, resulting positive $v_{\text{int}}$ in a large range of $i_{\text{bicone}}$. On the other hand, if $\theta_{\text{out}}$ is very wide (i.e., $\theta_{\text{out}}$=60\degree), we find positive $v_{\text{int}}$ values in a very small range of $i_{\text{bicone}}$ since it is more difficult to obscure the entire receding cone with a wider opening angle. In particular, if 
$\theta_{\text{out}}>i_{\text{dust}}$, outflows do not present positive $v_{\text{int}}$. In the case of $\sigma_{\text{int}}$, 
$\sigma_{\text{int}}$ increases at low $i_{\text{bicone}}$ if $\theta_{\text{out}}$ increases. This is due to the broadening of the LOSVD
with larger $\theta_{\text{out}}$. In contrast, for type 1 AGNs (i.e., high $i_{\text{bicone}}$), the trend becomes opposite. 

We also examine the effect of $\theta_{\text{in}}$ by removing the hollow region, i.e., $\theta_{\text{in}}$=0\degree\ (dashed lines in Figure \ref{fig:1doangle}). For both $v_{\text{int}}$ and $\sigma_{\text{int}}$, the difference between minimum and maximum values becomes slightly (a few percent) larger compared to the case of the hollow region, since the filled hollow regions from each cone make the LOSVD broader.  In other words, the effect of smaller $\theta_{\text{in}}$ is similar to that of larger $\theta_{\text{out}}$

\begin{figure*}
\centering
\includegraphics[width=1.0\textwidth]{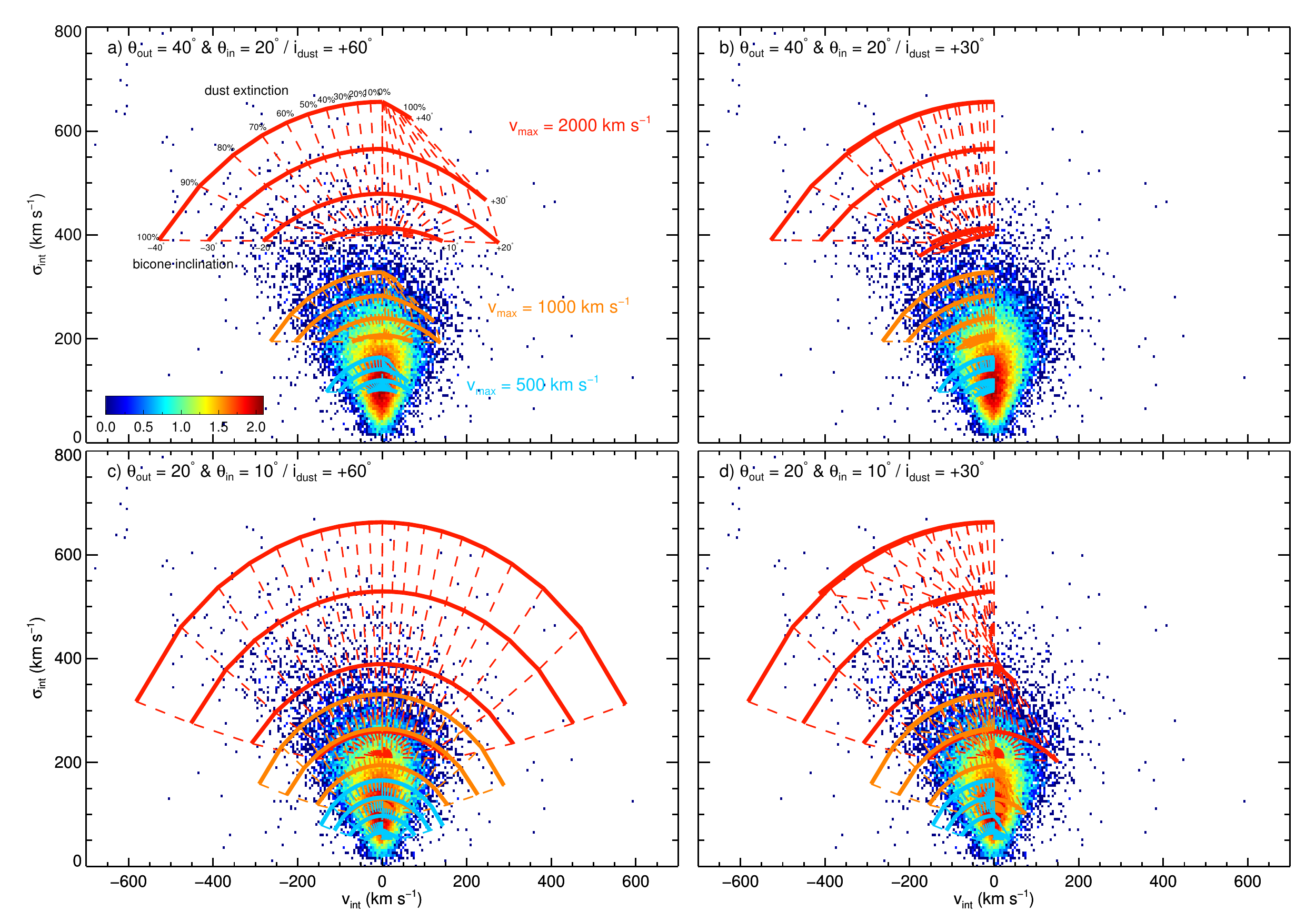}
\caption{Model grids constructed from the biconical outflow models (colored lines) as a demonstration, overlaid with the observed VVD distribution of $\sim$39,000 type 2 AGNs (squares). The model grids are based on a different set of the opening angles and dust extinction: a) $\theta_{\text{out}}$=40\degree, $\theta_{\text{in}}$=20\degree, and $i_{\text{bicone}}$=+60\degree, b) the same as a) but $i_{\text{bicone}}$=+30\degree, c) the same as a) but $\theta_{\text{out}}$=60\degree, $\theta_{\text{in}}$=30\degree, and d) the same as c) but $i_{\text{bicone}}$=+30\degree. The model grids are constructed with a different combination of nine different bicone inclinations (from $-$40\degree to $+$40\degree\ with 10\degree\ interval), 11 levels of dust extinction (from 0\% to 100\% with 10\% interval). Solid lines indicate the effect of dust extinction at fixed $i_{\text{bicone}}$, while dashed lines indicate the effect of $i_{\text{bicone}}$ at fixed dust extinction. For each panel, we show three different model grids with different $v_{\text{max}}$: from the top to bottom, $v_{\text{max}}$ is 2000 \kms\ (red lines), 1000 \kms\ (orange lines), and 500 \kms\ (blue lines), respectively. We label the value of bicone inclination or dust extinction at each end of the grid line on the model grid with $v_{\text{max}}$ = 2000 \kms\ in top-left panel (a).}
\label{fig:grid}
\end{figure*}

\section{Simulating the emission-line profile}
\label{sec:profile}
In this section, we construct a set of emission-line profiles based on the 3D bicone models to compare with the observed [\OIII] line profiles. Here we assume $\theta_{\text{out}}$=40\degree, and $\theta_{\text{in}}$=20\degree, and PA$_{\text{bicone}}$=PA$_{\text{dust}}$=0\degree, for simplicity. To focus on more realistic profiles of velocity and flux, we assume the linearly decreasing $v_{d}$ and the exponentially decreasing $f_{d}$ with $\tau=5$, which reduces the flux at the outskirts of bicone down to $\sim$1\% of the central value, to match with the observed [\OIII] flux distributions \citep[e.g.,][]{2013ApJS..209....1F}. We adopt five different values of $v_{\text{max}}$=[200, 400, 600, 800, 1000] \kms. The bicone inclinations are fixed at $-$30\degree, 0\degree, and 30\degree, respectively (top to bottom in Figure \ref{fig:profile}). The amplitude of the line profiles without dust extinction is normalized to one. To construct an emission-line profile for a given set of model parameters, first of all, we obtain an LOSVD for each LOS at ($x,y$) position and then combine them together over $x$ and $y$. We apply a Gaussian kernel $\sigma_k$ to the LOSVD with the velocity dispersion of 70 \kms, which is similar to the instrumental resolution of SDSS. 

First, we consider the simple case of no dust extinction (black lines in Figure \ref{fig:profile}). In this case, $v_{\text{int}}$ is always equal to zero due to the cancelation of approaching and receding gas. As $v_{\text{max}}$ increases, the line profiles become broader and a broad wing feature becomes more noticeable compared to a narrow core component in the line profile. 

Second, we examine the effects of the dust extinction on the line profiles with two different inclinations of the dust plane as $i_{\text{dust}}$=+45\degree\ or $i_{\text{dust}}=-$45\degree\ with a fixed dust extinction ($A$=90\%). Since the dust obscures a part of the bicone in both cases, the flux centroid of the line profile ($v_{\text{int}}$) offsets from zero velocity, showing asymmetric profiles when $i_{\text{bicone}}$=$\pm$30\degree\ (top and bottom). In such cases, |$v_{\text{int}}$| becomes larger as $v_{\text{max}}$, while the peak velocity of the line does not change much. In contrast, if $i_{\text{bicone}}$=0\degree, the LOSVD is symmetric because one of the cones is fully visible (i.e., not affected by the dust extinction), while the other cone is fully obscured. 

Third, we investigate the line profiles by varying the radial flux profile exponent $\tau$ and the bicone opening angle $\theta_{\text{out}}$. By using the linearly decreasing velocity profile with $v_{\text{max}}$=1000 \kms\ and $i_{\text{bicone}}$=$-$30\degree\ for simplicity, we examine the effect of five different exponent values of $\tau$=[3, 4, 5, 6, 7] in radial flux distribution (Figure \ref{fig:tau}). If the flux decreases more rapidly (i.e., increasing $\tau$), the narrow component becomes weaker compared to the broad wing feature, since the flux contribution from the outskirts becomes weaker. On the other hand, if the flux decreases more slowly (i.e., decreasing $\tau$), the narrow core component dominates over the broad wing feature due to the strong contribution of low velocity gas at the outskirts.

To examine the effect of the outer opening angle of the bicone $\theta_{\text{out}}$, we use three different $\theta_{\text{out}}$=[20\degree, 40\degree, 60\degree] with the inner opening angle as a half of $\theta_{\text{out}}$. We assume $v_{\text{max}}$=1000 \kms, $i_{\text{bicone}}=-$30\degree\ and 0\degree\ for simplicity (top and bottom in Figure \ref{fig:oa}, respectively). The line width becomes larger as $\theta_{\text{out}}$ increases regardless of dust extinction. If there is no dust extinction (black lines) and $i_{\text{bicone}}$=0\degree\ (bottom), the line profile becomes broader as $\theta_{\text{out}}$ becomes larger. Interestingly, if $\theta_{\text{out}} < |i_{\text{bicone}}|$ (top-left), the line shows a double-peak profile since the differences between positive and negative velocities become very large. If we add the dust plane (e.g., $i_{\text{dust}}$=+45\degree\ or $i_{\text{dust}}=-$45\degree), the dust obscures a part of each cone, reducing the chance of a double-peak profile.  

\section{Model grids for the observed [\OIII] kinematics}
\label{sec:grid}
In this section, we present model grids of $v_{\text{int}}$ and $\sigma_{\text{int}}$ with various combinations of geometrical and kinematical parameters (Figure \ref{fig:grid}, colored lines). Here we assume that the velocity is linearly decreasing and the flux is exponentially decreasing with $\tau$=5. For each model grid, we use nine different bicone inclinations (from $-$40\degree\ to $+$40\degree\ with 10\degree\ interval) and 11 levels of dust extinction (from 0\% to 100\% with 10\% interval) with three different $v_{\text{max}}$ (500 \kms, 1000 \kms, and 2000 \kms). Solid lines indicate the effect of dust extinction at fixed $i_{\text{bicone}}$, while dashed lines indicate the effect of $i_{\text{bicone}}$ at fixed dust extinction. We also present model grids based on the different sets of dust inclination (+30\degree, +60\degree) and bicone opening angles ([$\theta_{\text{out}}$, $\theta_{\text{in}}$] = [40\degree, 20\degree], [20\degree, 10\degree]). 

To compare the model grids with observations, we overlay the observed VVD distribution of the [\OIII] line measured from $\sim$39,000 type 2 AGNs (Paper I) in Figure \ref{fig:grid} (squares). The different color indicates the number of AGNs at each bin (5 \kms\ $\times$ 5 \kms) in a logarithmic scale, as indicated by the color bar in Figure \ref{fig:grid}a. 
 
The model grids clearly show the characteristics of both $v_{\text{int}}$ and $\sigma_{\text{int}}$ depending on the model parameters, as we found in the previous sections. 
For example, the model grid with higher $v_{\text{max}}$ has higher $\sigma_{\text{int}}$ and a larger range of $v_{\text{int}}$ (see Section \ref{sec:profile}). The model grid with $v_{\text{max}}$=2000 \kms\ covers $\sim$93\% (326 out of 351) of the observed AGNs with extreme kinematics, i.e., $\sigma_{\text{[O III]}} > $400 \kms, while a densely populated region in the observed VVD distribution around $\sigma_{\text{[O III]}} \sim$100 \kms\ is mostly covered by the model grid with $v_{\text{max}}$=500 \kms.  

The model grids with different $v_{\text{max}}$ well reproduce the range of the observed [\OIII] VVD distribution, suggesting that the biconical outflow models (with a range of $v_{\text{max}}$) can reproduce the observed VVD distribution. The inner angle of the V-shape of the VVD distribution depends on the amount of dust extinction, which is one of the primary parameters for determining $v_{\text{int}}$ (see Section \ref{sec:bincl}). If the dust extinction is higher, the angle of the V-shape is larger, and vice versa. 

At fixed $v_{\text{max}}$, the grids show that $\sigma_{\text{int}}$ depends on $i_{\text{bicone}}$ and dust extinction: $\sigma_{\text{int}}$ increases as $i_{\text{bicone}}$ increases, while $\sigma_{\text{int}}$ decreases as dust extinction increases. While the maximum $\sigma_{\text{int}}$ remains almost the same regardless of $\theta_{\text{out}}$, the minimum $\sigma_{\text{int}}$ becomes smaller as $\theta_{\text{out}}$ becomes smaller. Also, if $\theta_{\text{out}}$ is larger than $i_{\text{dust}}$ (Figure \ref{fig:grid}b), the model grids do not cover positive $v_{\text{int}}$ since there is no chance to produce positive $v_{\text{int}}$ (see Section \ref{sec:boa}). In contrast, if $\theta_{\text{out}}$ is small enough compared to $i_{\text{dust}}$ (Fig \ref{fig:grid}c), the grids span the full range of velocities from negative to positive values since the dust plane can obscure the entire approaching or receding cone in that geometry. As $i_{\text{dust}}$ becomes smaller, but still larger than $\theta_{\text{out}}$ (Fig \ref{fig:grid}d), the model grids cover the same range of negative $v_{\text{int}}$, but a smaller range of positive $v_{\text{int}}$ compared to the case with higher $i_{\text{dust}}$.

\begin{figure}
\centering
\includegraphics[width=0.49\textwidth]{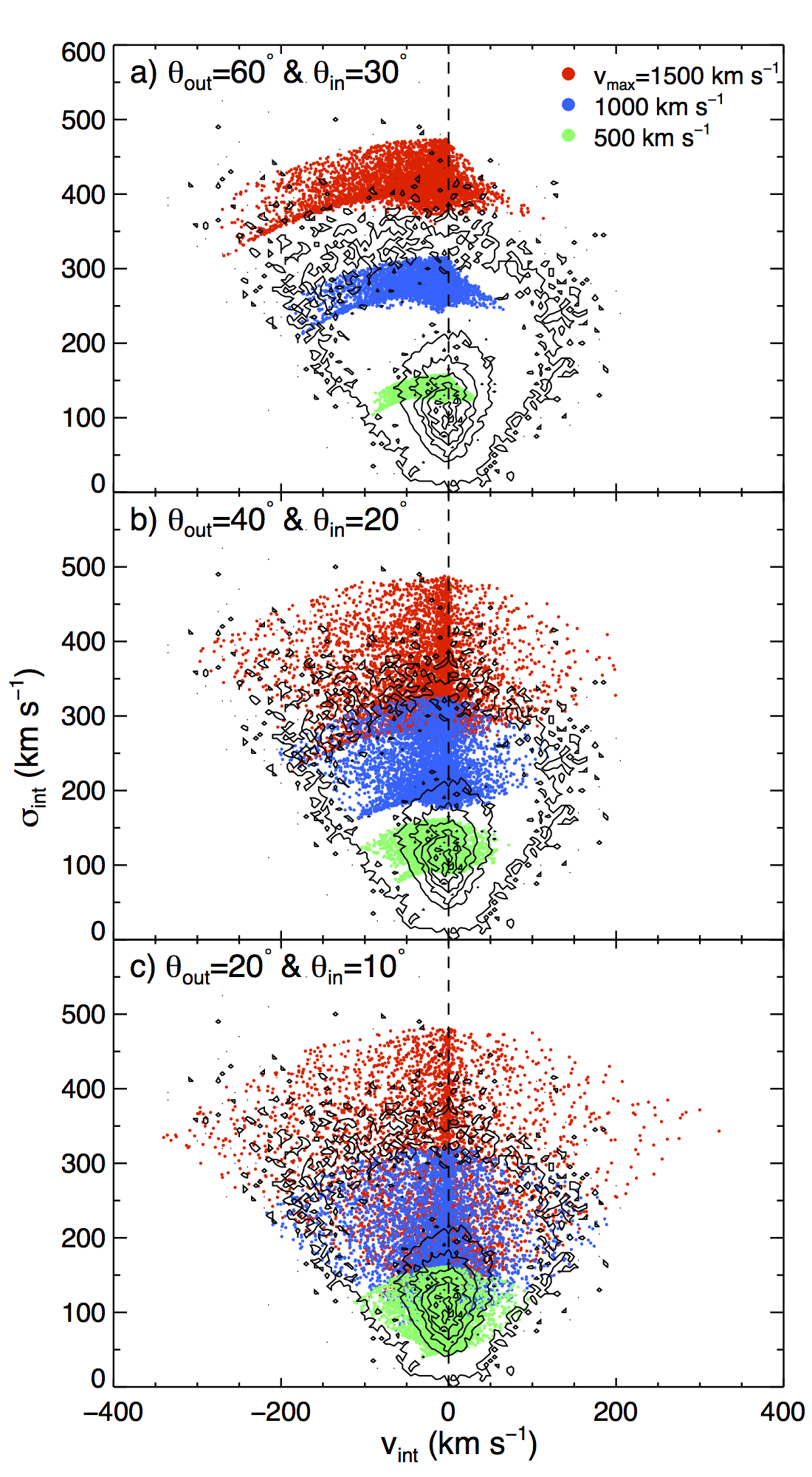}
\caption{Simulated velocity and velocity dispersion diagrams based on the bicone models with a different set of opening angles [$\theta_{\text{out}}$, $\theta_{\text{in}}$]: a) [60\degree, 30\degree], b) [40\degree, 20\degree], and c) [20\degree, 10\degree]. Each color represents the results from a different maximum velocity, i.e., $v_{\text{max}}$=[500 (green), 1000 (blue), 1500 (red)]. We assume that the velocity is linearly decreasing, and that the flux is exponentially decreasing with $\tau$=5. Contours represent the number density of the observed VVD for the [\OIII] line in type 2 AGNs with levels of [1, 20, 40, 60, 80, 100].}
\label{fig:vvd_mc_total}
\end{figure}

\section{Monte Carlo simulations for the observed [\OIII] kinematics}
\label{sec:mc}
\subsection{Simulating VVD distributions}
\label{sec:simvvd}
While the model grids in the previous section constrain the range of the LOS velocity and velocity dispersion of outflows in type 2 AGNs, 
Monte Carlo simulations can provide the probability of AGNs populating in a particular part of the VVD diagram, enabling us to investigate
the distribution of AGNs in the VVD diagram. In this section, we perform Monte Carlo simulations to investigate the intrinsic properties of gas outflows
in type 2 AGNs.

For simulations, we use a set of parameters as follows. We assume that outflow velocity is radially decreasing and flux is exponentially decreasing with $\tau$=5, as we used in the model grid constructions. We randomly choose the rotation angles of the bicone and the dust plane from a uniform distribution between 0\degree\ and 180\degree\ (see Figure \ref{fig:euler}). Similarly, dust extinction values are selected randomly from a uniform distribution ranging from 0\% to 100\%. We construct the VVD distribution, mainly using three different combinations of parameters, i.e., [$\theta_{\text{out}}$, $\theta_{\text{in}}$] = [20\degree, 10\degree], [40\degree, 20\degree], and [60\degree, 30\degree]. In the case of the intrinsic velocity, we use three representative velocities, i.e., $v_{\text{max}}$=[500, 1000, 1500] \kms. For each Monte Carlo simulation, we generate 10,000 AGNs for each set of parameters, in order to reproduce
the VVD distribution. 
   
Note that the uniform distribution of rotation angles of the bicone and the dust plane naturally produces a skewed distribution of both bicone inclination and dust inclination with respect to an observer (Equation (1) and (3)). For example, Monte Carlo simulations generate a larger number of AGNs with small bicone inclinations (i.e., type 2) than AGNs with high inclinations (i.e., type 1). If we divide type 1 and type 2 AGNs at the inclination angle $|i_{\text{bicone}}|=$40\degree, the average ratio of type 2 AGNs to type 1 AGNs is $\sim$3:1 as expected from the unification model. In the following, we present the Monte Carlo simulation results by restricting $|i_{\text{bicone}}|<$40\degree, following the expected inclination range of type 2 AGNs, as also adopted in the model grid construction in Section \ref{sec:grid}. We present the Monte Carlo simulation results with the dust extinction $A<$90\% (corresponding to lower than 2.5 magnitude), in order to match the angle of V-shape in the observed VVD distribution taken from Paper I (black contours in Figure \ref{fig:vvd_mc_total}). In other words, we rarely see AGNs with dust extinction $A>$90\% in the observed VVD diagram. Although it is difficult to characterize the level of dust extinction, our empirical constraint ($A<$90\%) based on the shape of the VVD distribution is consistent with other observations in that there is no such extreme dust extinction from galaxies in the local universe \citep[e.g.,][]{Holwerda:2005dl,Holwerda:2014dn}.  

In Figure \ref{fig:vvd_mc_total} we present the Monte Carlo simulation results compared to the distribution of the observed VVD distribution of $\sim$39,000 local type 2 AGNs 
from Paper I (contours in Figure \ref{fig:vvd_mc_total}). By comparing the simulation results with different opening angles, we find that for given intrinsic velocity $v_{\text{max}}$, the maximum $\sigma_{\text{int}}$ remains almost the same regardless of $\theta_{\text{out}}$, while the minimum $\sigma_{\text{int}}$ decreases if we use smaller $\theta_{\text{out}}$. This trend is also expected from the model grid calculations (see Figure \ref{fig:grid}). 

The Monte Carlo results clearly show a different density on a specific region in the VVD diagram depending on various model parameters. First, at fixed intrinsic velocity $v_{\text{max}}$, the region with lower $v_{\text{int}}$ and lower $\sigma_{\text{int}}$ in the VVD diagram has the highest density, reflecting the skewed distribution of $i_{\text{bicone}}$ and $i_{\text{dust}}$ (see, for example, the blue points in Figure \ref{fig:vvd_mc_total}). Second, the uniform distribution of dust extinction produces a high number density near the region at $v_{\text{int}}$= 0 \kms. This means that there are more AGNs with no, or very weak, velocity shifts with respect to systemic velocity than AGNs with high velocity shifts. Third, the number of negative $v_{\text{int}}$ is always larger than the number of positive $v_{\text{int}}$ as we expect from the parameter test (see Section \ref{sec:test}). Interestingly, a wider opening angle produces a larger number ratio between negative $v_{\text{int}}$ and positive $v_{\text{int}}$, because the bicone with a larger opening angle has a smaller chance of producing positive $v_{\text{int}}$ (see also Section \ref{sec:boa}). In the Monte Carlo simulations, the number ratio of negative $v_{\text{int}}$ to positive $v_{\text{int}}$ ($N_n/N_p$) is [$\sim$1.6, $\sim$2.6, $\sim$7.2] for $\theta_{\text{out}}$=[20\degree, 40\degree, 60\degree], respectively.  

We find that the intrinsic velocity of gas outflows ranges from several hundred to a few thousand \kms\ based on the comparison of the Monte Carlo
simulations with the observed VVD distribution. 
For example, the Monte Carlo simulations with $v_{\text{max}}$=500 and 1000 \kms\ and $\theta_{\text{out}}$=20\degree\ produce the VVD range of AGNs with $\sigma_{\text{int}}$ from $\sim$50 to 300 \kms, which constitute the majority ($\sim$94\%) of the type 2 AGNs (green and blue points in Figure \ref{fig:vvd_mc_total}),
while the model results with $v_{\text{max}}$=1500 \kms\ and $\theta_{\text{out}}$=60\degree\ can reproduce the VVD of AGNs with higher 
velocity dispersion (i.e., $\sigma_{\text{[O III]}}$>$\sim$350 \kms; red points).
Note that it is difficult to constrain the distribution of intrinsic velocity for AGNs with weak outflows since the signature of low velocity and velocity dispersion may be washed out by the virial motion due to the host galaxy's gravitational potential. Hence, we need to investigate the intrinsic velocity of gas outflows by focusing on AGNs with strong outflow kinematics (i.e. high velocity or velocity dispersion).
The intrinsic gas outflow velocities constrained by comparing the Monte Carlo simulation results with the observed VVD distribution are consistent with those of other studies in the literature. For example, \citet{2013ApJS..209....1F} found a range of $v_{\text{max}}$ from 130 to 2000 \kms\ with the mean value of 840$\pm$530 \kms\ based on bicone models for 17 local Seyfert galaxies.   

\subsection{Constraining the opening angle}
\label{sec:oa}
The Monte Carlo simulation results show that models with a higher bicone opening angle $\theta_{\text{out}}$ produce a larger ratio of the number of AGNs with negative velocity ($N_n$) to the number of AGNs with positive velocity ($N_p$) (see Figure \ref{fig:vvd_mc_total}). To compare this trend with observations, we calculate the $N_n/N_p$ ratio of 
the SDSS type 2 AGNs in our previous study (Paper I), in which we measured the [\OIII] velocity of each AGN with a mean measurement uncertainty of
$\sim$27 \kms. If we select $\sim$20,000 AGNs with reliable velocity measurements (i.e., |$v_{\text{[O III]}}$| >1$\sigma$ uncertainty),
after accounting for the measurement uncertainties, we find a clear trend that the $N_n/N_p$ ratio increases with [\OIII] velocity dispersion or luminosity 
(Table \ref{tbl-2}). These results imply that more energetic AGNs with a larger velocity dispersion and luminosity have, on average, higher $\theta_{\text{out}}$. For example, the $N_n/N_p$ ratio is $\sim$4.5 for AGNs with $\sigma_{\text{[O III]}}$ larger than 400 \kms, while the number ratio dramatically decreases to $\sim$1.1 for AGNs with a lower velocity dispersion of 100$-$200 \kms. 
Although [\OIII] velocity is more difficult to measure for AGNs with weaker [\OIII] lines
due to the measurement uncertainty and the effect of the host galaxy's virial motion, the increasing trend of the $N_n/N_p$ ratio is significant even if we limit the sample of AGNs with a high velocity dispersion (i.e., $\sigma_{\text{[O III]}}$>200 \kms). If we further restrict the sample by focusing on AGNs with more robust velocity measurements (i.e., |$v_{\text{[O III]}}$| >3$\sigma$ uncertainty), the trend becomes even stronger than the former case. 

By comparing the simulated $N_n/N_p$ ratios with the observed ones, we constrain $\theta_{\text{out}}=\sim$20\degree\ for AGNs with low velocity dispersion (i.e., $\sigma_{\text{[O III]}}<$300 \kms), $\theta_{\text{out}}=\sim$40\degree\ for the AGNs with high velocity dispersions (i.e., 300 \kms $<\sigma_{\text{[O III]}}<$400 \kms), and $\theta_{\text{out}}=\sim$50-60\degree\ for the AGNs with extreme velocity dispersions (i.e., $\sigma_{\text{[O III]}}>$400 \kms).

\begin{table}
\center
\caption{$N_n/N_p$ Ratios from Observation \label{tbl-2}}
\begin{tabular}{cccc}
\tableline
\tableline
$\sigma_{\text{[O III]}}$ Range& $N_n$ & $N_p$ & $N_n/N_p$ \\
(1) & (2) & (3) & (4) \\
\tableline
0 -- 100 & 1510 & 1622 & 0.9 \\
100 -- 200  & 6073 & 5314 & 1.1\\
200 -- 300 & 2346 & 1724 & 1.4 \\
300 -- 400 & 620 & 252 & 2.5 \\
$>$ 400  & 243 & 54 & 4.5 \\
\tableline
\end{tabular}
\tablecomments{(1) range of [\OIII] velocity dispersion (\kms); (2) number of AGNs with negative $v_{\text{[O III]}}$, where |$v_{\text{[O III]}}$| is larger than the measurement uncertainty; (3) number of AGNs with positive $v_{\text{[O III]}}$, where |$v_{\text{[O III]}}$| is larger than the measurement uncertainty; (4) number ratio of $N_n/N_p$.}
\end{table}

\section{Discussion}
\label{sec:discussion}

\subsection{Intrinsic Outflow Properties of Type 2 AGNs}
\label{sec:prop}
We find that our 3D gas outflow models successfully reproduce a number of essential characteristics 
of the observed [\OIII] gas kinematics and the VVD distribution of local type 2 AGNs, allowing us to constrain the intrinsic physical properties of outflows.
Previous studies in the literature had investigated the intrinsic outflow properties of local Seyfert galaxies by combining bicone models 
with high-resolution photometry and spectroscopy data \citep[e.g.][]{2011ApJ...739...69M,2013ApJS..209....1F}.
The physical parameters of the gas outflows, e.g., intrinsic velocity, inclination of the bicone, and opening angle, were derived 
by fitting the observed velocity distribution of gas outflows in 1D or 2D, while spatially revolved velocity dispersions of gas outflows
were not significantly used in the analysis. In addition, while the observed velocity distribution can be strongly affected by dust extinction, 
these studies did not fully investigate the effect of dust extinction in the observed gas kinematics. 

In contrast, we used the combined VVD diagram in constraining the kinematics of gas outflows, and investigated 
the effect of dust extinction in detail. By examining the full range of physical parameters of gas outflows, including the radial distribution of 
velocity and flux in gas outflows, the inclination of the bicone and dust plane with respect to an observer, and the opening angle, 
we were able to understand how each parameter of gas outflows affects the LOS observations of gas kinematics. 
For the first time, we used Monte Carlo simulations compared to the observed kinematics of a large sample of type 2 AGNs to constrain the intrinsic properties of ionized gas outflows. 

The bicone opening angle seems to increase with the intrinsic velocity (Section \ref{sec:oa}). This trend is consistent with the expectation from the simple model of biconical outflow structures proposed by \citet{2000ApJ...545...63E}, who suggested that funnel-shaped outflows driven by accretion-disk winds explain the observed features in different physical scales from the inner broad absorption line regions to the outer NLR. In contrast, other studies reported no correlation between the opening angle and the intrinsic velocity of the outflows in the NLR \citep{2014ApJ...785...25F}, or an anti-correlation between opening angel and intrinsic velocity of outflows in the coronal-line region \citep{2011ApJ...739...69M}. 
Since our finding is based on the comparison of the spatially integrated kinematics with Monte Carlo simulations, it is difficult to directly
compare our results with the previous studies, which were mainly based on the spatially resolved measurements of gas kinematics.
Also, our constraint is mainly from the statistical analysis using a large sample, while the previous studies used relatively small samples. 
To better understand the relation between opening angle and outflow velocity, it is necessary to further investigate the intrinsic properties of gas outflows
by applying, for example, our 3D models to spatially resolved data. 

Outflows are of importance to understand the connection of AGNs with their host galaxies since outflows can be a channel for AGN feedback.
In observational studies, the [\OIII] line width, e.g., $\sigma_{\text{[O III]}}$ or $W_{80}$ (line width containing 80\% of the line flux), is widely used to infer the bulk outflow velocity, i.e., $v_{\text{bulk}}\propto \sigma_{\text{[O III]}}\propto W_{80}$ \citep[e.g.,][]{2012MNRAS.425L..66M,2013MNRAS.436.2576L,2015A&A...580A.102C}. Then, $v_{\text{bulk}}$ is used to calculate the physical properties of outflows, e.g., energy injection rate as $\dot{E}\propto \dot{M} v_{\text{bulk}}^2$, where $M$ is the mass of gas, and $\dot{M}$ is the mass outflow rate, which is proportional to bulk velocity, i.e., $\dot{M}\propto v_{\text{bulk}}$. Thus, measuring the bulk velocity is important; however, since the measured [\OIII] line width, i.e., $\sigma_{\text{[O III]}}$, is used as a proxy for the bulk velocity, 
it is likely that the bulk velocity is significantly underestimated since the line width is affected by the bicone inclination and the dust extinction (see section \ref{sec:bincl} and \ref{sec:grid}). 

\begin{figure}
\centering
\includegraphics[width=0.49\textwidth]{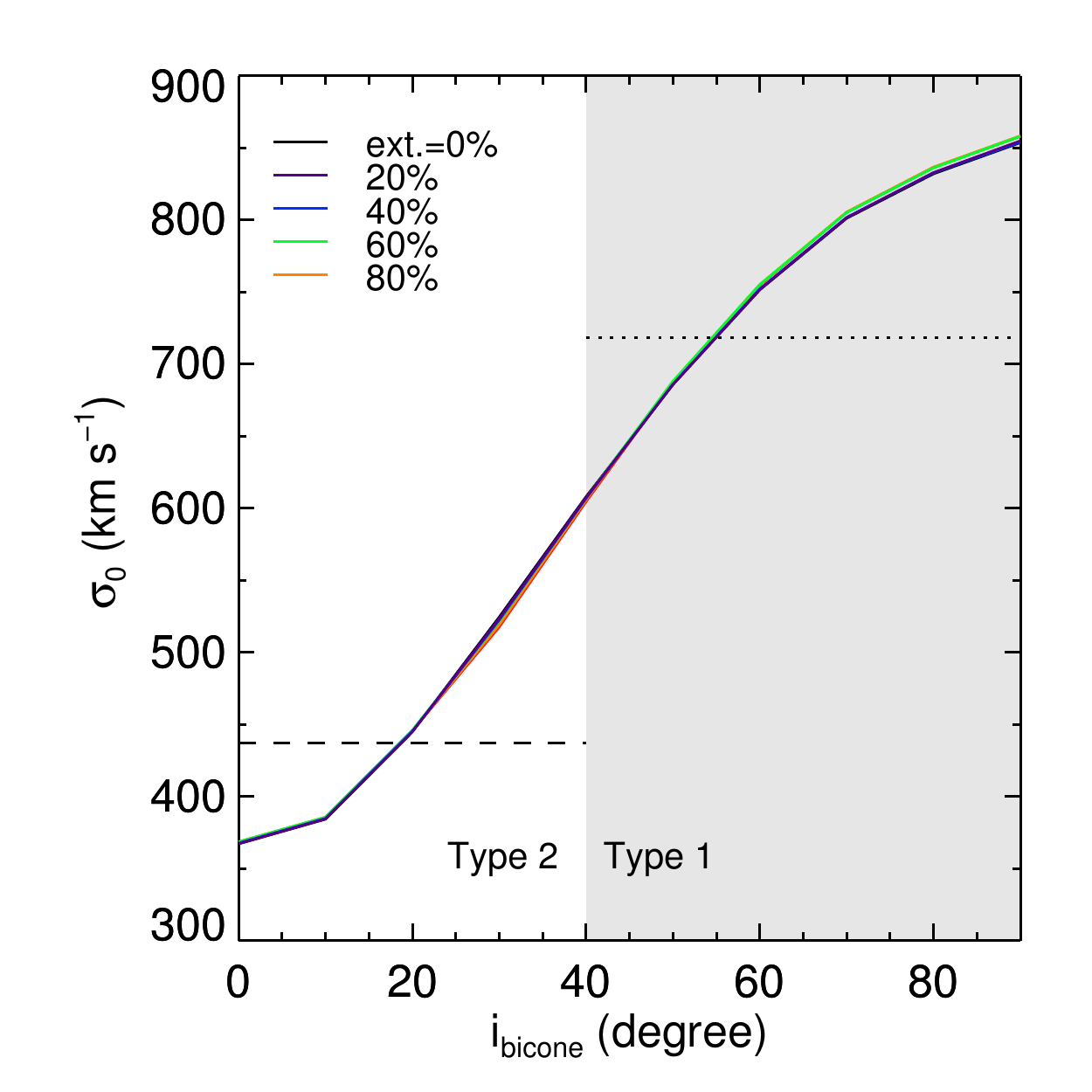}
\caption{Quadrature sum of velocity and velocity dispersion 
($\sigma_0$ = $[\sigma_{\text{int}}^2 + v_{\text{int}}^2]^{0.5}$)
as a function of bicone inclination. Different colored lines represent various amounts of dust extinction, which are negligible. The shaded and unshaded regions represent the range of the bicone inclinations for type 1 and 2 AGNs, respectively. The dotted and dashed line represents the mean $\sigma_0$ for type 1 and 2 AGNs, respectively, after considering the probability of $i_{\text{bicone}}$ for each type of AGN. For these simulations, we assume the constant velocity as $v_{\text{max}}$=1000 \kms, the constant radial flux, $\theta_{\text{out}}$=40\degree, $\theta_{\text{in}}$=20\degree, and $i_{\text{dust}}$=60\degree.}
\label{fig:sigma_0}
\end{figure}

For example, if the amount of dust extinction is as large as $\sim$90\%,  $\sigma_{\text{int}}$ decreases by $\sim$30\% for type 2 AGNs if the inclination changes from 0\degree\ to 40\degree\ (see Section \ref{sec:simvvd}). Hence, the estimated $\dot{M}\propto v_{\text{bulk}}$ can be underestimated by $\sim$30\%, while the energy injection rate $\dot{E}\propto \dot{M} v_{\text{bulk}}^2 \propto v_{\text{bulk}}^3$ can be underestimated by $\sim$70\%. 
Since dust extinction decreases $\sigma_{\text{int}}$, but increases $v_{\text{int}}$, we find a remedy for the dust extinction effect by combining 
$\sigma_{\text{int}}$ and $v_{\text{int}}$. The quadrature sum of $\sigma_{\text{int}}$ and $v_{\text{int}}$ is a good approximation of the dust-corrected
velocity dispersion ($\sigma_{\text{0}}$), which we calculate from our model by excluding the dust extinction effect (see Figure \ref{fig:sigma_0}).
In practice, $v_{\text{int}}$ itself is difficult to measure in many cases due to the lack of systemic velocity measurements. 

A more severe limitation of estimating the bulk velocity of gas outflows is due to the unknown bicone inclination.
Since the bicone inclination ranges from 0\degree\ to 40\degree\ for type 2 AGNs, $\sigma_{\text{int}}$ can significantly increase with the inclination angle.
For example, if we use $\theta_{\text{out}}$=40\degree\ for the model, $\sigma_{\text{int}}$  increases by $\sim$50\% from the case of $i_{\text{bicone}} = $0\degree . 
If we use a smaller opening angle, i.e., $\theta_{\text{out}}$=20\degree, $\sigma_{\text{int}}$ increases by a factor of $\sim$3.
Since the estimated bulk velocity and $\dot{M}$ strongly depend on the inclination of the bicone, the inclination effect should be carefully considered
in estimating feedback energy. 
Note that the effect of bicone inclination is much more significant in type 1 AGNs than in type 2 AGNs, since type 1 AGNs are expected to have more inclined bicone compared to type 2 AGNs. In Figure 14 we present the quadrature sum of $\sigma_{\text{int}}$  and $v_{\text{int}}$ (i.e., $\sigma_0$) as 
a function of the bicone inclination angle. Based on a Monte Carlo simulation with a uniform distribution of bicone orientation angles and $\theta_{\text{out}}$=40\degree, we find that $\sigma_0$ of type 1 AGNs is larger by an average factor of $\sim$1.6 than that of type 2 AGNs. This systemic difference of [\OIII] line width between type 1 and type 2 AGNs can be directly tested using a combined sample of type 1 and type 2 AGNs.

\subsection{Limitations of the Biconical Outflow Models}
While our 3D outflow models well reproduce the observed characteristics of the [\OIII] VVD distribution, it has limitations. For example, in the models, we do not include the effect of the host galaxy's gravitational potential \citep[e.g.,][Paper I]{2014ApJ...786....3S,2016ApJ...819..148K}. Such a gravitational effect additionally increases $\sigma_{\text{int}}$ of ionized gas, and also affects $v_{\text{int}}$. Nevertheless, the stellar velocity dispersion, which represents the gravitational potential, is relatively small (100$-$200 \kms) for typical galaxies in the sample of type 2 AGNs (see Paper I).
Thus, the additional kinematic component does not significantly change the observed kinematics for AGNs with powerful outflows (e.g., $\sigma_{\text{[O III]}}$>200 \kms).

Another potential contamination of the gas kinematics is due to gas inflows as observed in the local Seyfert galaxies \citep[e.g.,][]{2009A&A...500.1287S,2014MNRAS.438.3322S}. If the inflow is present, the observed features will be similar to those of outflow, but with an opposite sign of $v_{\text{int}}$. The approaching, infalling gas has a larger chance of being obscured than the receding, infalling gas, due to the dust in the stellar disk. Thus, AGNs with positive velocity are more likely to be
detected than AGNs with negative velocity. However, this expectation is not consistent with the observed VVD distribution of type 2 AGNs, suggesting that inflow does not seem strong among local type 2 AGNs. 

We note that we use a simple dust model with wavelength-independent opacity. Thus, these models are expected to be applied
to a single emission line for simulating gas kinematics. For comparison with more detailed kinematic studies based on multiple emission-lines,
it is necessary to use our models with different extinction values for each emission line since the dust extinction is higher at shorter wavelengths.  

\section{Summary and Conclusion}
\label{sec:summary}
We constructed 3D models of biconical outflows combined with a thin dust plane in order to investigate the physical properties of gas outflows.
After examining the effect of each model parameter on the simulated velocity and velocity dispersion of gas outflow, 
we presented the simulated emission-line profiles, and compared model grids and Monte Carlo simulations with the observed [\OIII] VVD diagram of local type 2 AGNs taken from Paper I. We summarize our findings as follows.
\medskip

1. The primary drivers of the spatially-integrated [\OIII] velocity and velocity dispersion are the intrinsic velocity, the bicone inclination, and the amount of dust extinction. Velocity dispersion $\sigma_{\text{int}}$ increases as the intrinsic velocity or the bicone inclination increases, while velocity $v_{\text{int}}$ (velocity shift with respect to systemic velocity) increases as the amount of dust extinction increases. 
\medskip
 
2. The simulated emission-line profiles based on the LOSVD
well reproduce a narrow core and a broad wing components in the observed [\OIII] line profiles of type 2 AGNs.  
\medskip

3. By comparing the model grids with the observed VVD diagram of 39,000 type 2 AGNs, we constrain the intrinsic velocity of gas outflows, ranging from $\sim$500 \kms\ to 1000 \kms\ for the majority of AGNs, and up to $\sim$1500--2000 \kms\ for extreme cases.
\medskip

4. The observed increase of the number ratio of AGNs with negative [\OIII] velocity to AGNs with positive [\OIII] velocity 
is well reproduced by Monte Carlo simulations,
suggesting that AGNs with higher intrinsic velocities tend to have wider opening angles.
\medskip

5. Monte Carlo simulations based on our 3D outflow models well reproduce the observed [\OIII] VVD distribution of local type 2 AGNs,
demonstrating the potential of the model for studying the physical properties of gas outflows. 
\medskip

Although various physical parameters of outflows, i.e., intrinsic velocity and inclination of the bicone, cannot be directly measured,
our models can provide valuable constraints on these parameters.
Our 3D models are applicable to spatially resolved kinematics,
for investigating the spatial distribution of gas kinematics. In the future, we will investigate the gas outflows in 2D by comparing
the results from integral-field spectroscopy with our model simulations. 

\acknowledgments
We thank the anonymous referee for his/her valuable comments on the manuscript. The work of H.J.B. was supported by NRF (National Research Foundation of Korea) Grant funded by the Korean Government (NRF-2010-Fostering Core Leaders of the Future Basic Science Program). J.H.W. acknowledges the support by the National Research Foundation of Korea  grant funded by the Korea government (No. 2016R1A2B3011457).

\end{document}